\documentclass[usenatbib,twocolumn]{mn2e}
\usepackage{epsfig}
\usepackage{amsmath}
\usepackage{lscape}
\usepackage{longtable}
\usepackage{anysize}
\usepackage{multirow}
\def\beq{\begin{equation}}
\def\eeq{\end{equation}}
\def\bey{\begin{eqnarray}}
\def\eey{\end{eqnarray}}

\def\Msun{\,{\rm M_\odot}}
\def\Msunh{\, h^{-1}{\rm M_\odot}}

\def\gs{\mathrel{\raise1.16pt\hbox{$>$}\kern-7.0pt
\lower3.06pt\hbox{{$\scriptstyle \sim$}}}}
\def\ls{\mathrel{\raise1.16pt\hbox{$<$}\kern-7.0pt
\lower3.06pt\hbox{{$\scriptstyle \sim$}}}}
\def\gtsima{$\; \buildrel > \over \sim \;$}
\def\ltsima{$\; \buildrel < \over \sim \;$}
\def\prosima{$\; \buildrel \propto \over \sim \;$}
\def\gsim{\lower.5ex\hbox{\gtsima}}
\def\lsim{\lower.5ex\hbox{\ltsima}}
\def\simgt{\lower.5ex\hbox{\gtsima}}
\def\simlt{\lower.5ex\hbox{\ltsima}}
\def\simpr{\lower.5ex\hbox{\prosima}}
\def\la{\lsim}
\def\ga{\gsim}

\usepackage{color}

\marginsize{1.0in}{1.0in}{1.0in}{1.0in}

\begin{document}

\title[]
{Galaxy Ecosystems: gas contents, inflows and outflows}
\author[Zhankui Lu et al.]
   {\parbox[t]{\textwidth}{
       Zhankui Lu$^{1}$\thanks{E-mail: lv@astro.umass.edu},
       H.J. Mo$^{1}$,
       Yu Lu$^{2}$
}\\
$^1$Department of Astronomy, University of Massachusetts, Amherst MA 01003-9305, USA\\
$^2$Kavli Institute for Particle Astrophysics and Cosmology, Stanford, CA 94309, USA}

\date{Accepted ........ Received .......; in original form ......}
\pubyear{2014}

\maketitle 

\label{firstpage}

\begin{abstract}
We use a set of observational data for galaxy cold gas mass fraction and gas phase metallicity
to constrain the content, inflow and outflow of gas in central galaxies hosted by halos with masses 
between $10^{11} M_{\odot}$ to $10^{12} M_{\odot}$. 
The gas contents in high redshift galaxies are obtained 
by combining the empirical star formation histories  
and star formation models that  relate star formation rate with the cold gas mass in galaxies. 
We find that the total baryon mass in low-mass galaxies is always much 
less than the universal baryon mass fraction since $z = 2$, 
regardless of  star formation model adopted. 
The data for the evolution of the gas phase metallicity require 
net metal outflow at $z\la 2$, and the metal loading factor is 
constrained to be about 0.01, or about 60\% of the metal yield. 
Based on the assumption that galactic outflow is 
more enriched in metal than both the interstellar medium and
the material ejected at earlier epochs, we are able to put stringent  
constraints on the upper limits for both the net accretion rate and 
the net mass outflow rate. The upper limits strongly suggest that 
the evolution of the gas phase metallicity and gas mass fraction 
for low-mass galaxies at $z < 2$ is not compatible with strong outflow.
We speculate that the low star formation efficiency of low-mass 
galaxies is owing to some preventative processes that prevent 
gas from accreting into galaxies in the first place.
\end{abstract}



\section{Introduction}
\label{intro}

During the past 10 years, great progress has been made in 
establishing the connection between galaxies and dark
matter halos with the use of various statistical methods
\citep[e.g.][]{Yang03, vdBosch03, Conroy06, Moster10, 
Behroozi10, Yang12}. In particular, empirical models have been
developed to describe the star formation and stellar mass
assembly histories of galaxies in dark matter halos at different
redshifts \citep[e.g.][]{Conroy09, Behroozi13, Yang13, 
Bethermin13, LZ14a,LZ14b}. The results obtained all 
show that star formation is the 
most efficient in $\sim 10^{12}\Msunh$ halos over a 
large range of redshift, and that the efficiency drops rapidly 
towards both the higher and lower mass ends. 

The physics that regulates star formation in galaxies  
has been one of the main research topics in the field 
of galaxy formation and evolution. The processes that can 
affect star formation are generally 
divided into three categories: gas inflow, outflow, 
and star formation. The low star formation efficiency can either 
be caused by a reduced gas inflow, a strong gas outflow driven 
by some feedback processes,  or the distribution, thermal and 
chemical states of the cold gas disk, but how the processes
work in detail is still unclear. For high-mass galaxies, 
the quenching of star formation is believed to stem from the 
suppression of gas inflow into the galaxies by processes, 
such as AGN heating \citep[e.g.][]{Croton06}, that can heat 
the gas supply. 

For less massive galaxies, one popular scenario is strong gas outflow driven 
by supernova (SN) explosions  and radiation pressure from
massive stars \citep[e.g.][]{Dekel86, Oppenheimer08}.
Because of the relatively shallow potential wells associated with 
low-mass galaxies,  outflows may drive gas out of their host halos, 
reducing the gas supply for star formation.
Preventative scenarios have also been proposed 
for low-mass galaxies. For example, using a simple analytic model
and observational constraints, \citet{Bouche10} suggests that
gas accretion must be suppressed if the halo mass is $<10^{11}\Msun$.
The physical mechanism is uncertain.
Heating by the UV background \citep{Ikeuchi86, Rees86}. 
is found to be effective only in halos with masses
below $\sim 10^{10}\Msunh$ \citep[e.g.][]{Gnedin00}. 
For halos with mass $\sim 10^{11}\Msunh$ 
other heating sources have been suggested, 
such as gravitational pancaking \citep{Mo05}, 
blazar heating \citep{Chang11}, and galactic winds 
\citep[e.g.][]{Mo02,Mo04,vdVoort11}.

It is also possible that the low star formation efficiency
is caused by a low efficiency of converting cold gas into stars. 
Indeed, as shown by \citet{Krumholz12} using the 
metallicity-regualted star formation model developed in
\citet{Krumholz08} and \citet{Krumholz09}, 
the star formation in very low-mass galaxies can be
completely shut off owing to their low metallicities.

To distinguish between the different scenarios, one ultimately 
needs direct observational constraints on inflow, outflow and the 
gas distribution in high redshift galaxies. In the absence  
of such direct observational data at the moment,  
observational measurements such as the metallicity of the 
interstellar medium \citep{Tremonti04, Erb06, Kewley08}, 
have been used to constrain gas flows and star formation in galaxies
\citep{Dalcanton07, Erb08, Peeples11, Lilly13, Zahid14}.
For instance, using a simple analytic chemical evolution model, 
\citet{Erb08} constrained the outflow and argued that the mass 
loading factor of the outflow should be about unity in order to match the 
mass-metallicity relation. Using a more sophisticated 
chemical evolution model that takes into account 
inflow/outflow of the gas and star formation in the ISM, 
\citet{Lilly13} were able to infer, from a set of simple but 
plausible assumptions, how the mass loading factor of the 
outflow depends on the mass of host galaxies

In this paper, we construct a model for the galactic 
ecosystem, which includes the gas content (both atomic 
and molecular), inflow and outflow of gas and the enrichment of 
metals. Using current observational data on the gas mass fraction 
of local galaxies \citep{Peeples11, Papastergis12} and the evolution 
of gas phase metallicity-stellar mass relation 
\citep{Maiolino08}, together with the empirically 
constrained star formation histories obtained by \citet{LZ14a}, 
we infer how gas inflow and outflow regulate star formation.  
The paper is organized as follows. 
The basic equations that govern the evolution 
of different components of a galaxy are described 
in \S\ref{model}. 
The observational constraints adopted
are presented in \S\ref{constraints}.
In \S\ref{gas_content},  we use two optional star 
formation laws (models) to infer the cold gas mass for galaxies 
with different masses and  at different redshifts.  
In \S\ref{io}, we constrain the mass and metal exchange between
galaxies and their environments to shed light on gas 
inflow and outflow, using information about 
the major components of galaxies, such as 
dark halos, stars, cold gas and metals. Finally 
we summarize our conclusions and discuss 
their implications for physically motivated galaxy 
evolution models in \S\ref{conclusion}.

Throughout the paper, we use a $\Lambda$CDM cosmology with
$\Omega_{\rm m,0}=0.273$, $\Omega_{\Lambda,0}=0.727$, 
$\Omega_{\rm b,0}=0.0455$, $h=0.704$, $n=0.967$ and $\sigma_{8}=0.811$.  
This set of parameters is from the seven year WMAP observations \citep{Komatsu11}.
Unless stated otherwise, we adopt the stellar population
synthesis model of \citet{Bruzual03} and a 
Chabrier IMF \citet{Chabrier03}.
Solar metallicity is defined as $Z_{\odot} = 0.0142$ 
in terms of the total metal mass fraction, and 
as $Z_{\odot, \rm O} = 0.0056$ in terms of  the oxygen mass fraction 
\citep{Asplund09}.


\section{Galaxy Ecology}
\label{model}

\subsection{The basic equations}

The ecosystem of a galaxy consists of stellar mass ($M_\star$), 
cold gas mass ($M_{\rm g}$), and metal mass ($M_{\rm Z}$).
The evolution of these components  
in a galaxy can be described by the following set of equations:
\begin{eqnarray}
 \label{eq:star}
 \frac{{\rm d} M_{\star}}{{\rm d}t} & = & (1-R)\Psi(M_{\rm g},\, R_{\rm g},\, Z)\,;  \\
 \label{eq:gas:01}
 \frac{{\rm d} M_{\rm g}}{{\rm d}t} & = & 
      \epsilon_{\rm acc} f_{\rm b}\dot{M}_{\rm h} - \dot{M}_{\rm w} +\dot{M}_{\rm r} - (1-R)\Psi \,;\\
 \label{eq:metal:01}
 \frac{{\rm d} M_{\rm Z}}{{\rm d}t} & = & 
      \epsilon_{\rm acc} f_{\rm b}\dot{M}_{\rm h}Z_{\rm IGM}  \\ \nonumber
    && - \dot{M}_{\rm w}Z_{\rm w} + \dot{M}_{\rm r}Z_{\rm r} \\ \nonumber
    && - (1-R)\Psi Z  + y \Psi \,.
\end{eqnarray}
In this set, Eq.\,(\ref{eq:star}) specifies the change in stellar mass, 
with $\Psi$ being the star formation rate and $R$ being the return 
mass fraction of evolved stars. Here we make instanteneous recycling
approximation, which is reasonable in the redshift range we consider
($0\le z\lsim 2$) because the lifetime of the stars that contribute
most of the recycling is short compared to the Hubble time.
The star formation rate 
depends on the mass ($M_{\rm g}$), distribution (characterized 
by the size of the cold gas disk, $R_{\rm g}$), metallicity ($Z$),
and perhaps other properties of the interstellar medium (ISM), 
as specified by a star formation law.

Eq.\,(\ref{eq:gas:01}) describes the evolution in cold gas mass.
The first term on the right hand side is the inflow rate of pristine gas, 
written in terms of mass accretion rate of the host dark halo,
$\dot{M}_{\rm h}$ (see \S\ref{model:hah})
multiplied by the universal baryon mass fraction, 
and a gas accretion efficiency,  $\epsilon_{\rm acc}$.
In normal circumstances, the efficiency factor $\epsilon_{\rm acc}\le1$,  
and its value may depend on halo mass and redshift.
This efficiency may be affected by a variety of physical processes.
For instance, if a halo is embedded in a preheated gas, the accretion 
into the halo may be reduced, making $\epsilon_{\rm acc} < 1$ \citep{LY07}.
It is also possible that the halo can accrete gas at a rate 
of $f_{\rm b}\dot{M}_{\rm h}$, but that certain heating sources 
such as ``radio-mode" AGN feedback in massive halos \citep{Croton06}
or photoionization heating by local sources \citep{Cantalupo10, Kannan14},
can prevent the coronal gas from cooling, making $\epsilon_{\rm acc}<1$. 
$\dot{M}_{\rm w}$ on the right hand of
Eq.\,(\ref{eq:gas:01}) is gas outflow,
and $\dot{M}_{\rm r}$ is the re-accretion rate of the gas mass that 
has been ejected at earlier times. Finally, the last term on the right hand side 
is the cold gas consumption rate of star formation.   
  
Eq.\,(\ref{eq:metal:01}) describes the chemical evolution.
$y$ is the intrinsic metal yield from stars. 
The metal yield is assumed to be instanteneous, which is
a good approximation, because we only consider
oxygen produced by short-lived massive stars (\S\ref{model:yield}).
$Z_{\rm IGM}$, $Z$,  $Z_{\rm w}$, and $Z_{\rm r}$ are
the metallicities of the intergalactic medium (IGM), 
ISM, wind, and the re-accreted material, respectively.
Note that we distinguish between the accretion of the pristine gas from the IGM 
and the re-accretion of the recycled wind material from 
the galaxy.  In general, the metallicity of the wind 
and recycled material is not necessarily equal to that of the ISM; 
for example supernova ejecta and stellar wind may
directly carry away metals 
\citep{MacLow99}, giving $Z_{\rm w} \ge Z$.

Note that this set of equations is only valid for galaxies
with no satellites of comparable masses. 
Otherwise, the central galaxies may obtain a significant amount 
of metals by accreting the enriched hot gas of the satellites 
after halo merger and the ISM once galaxy mergers occur.
In this paper we focus only on galaxies with their halo mass
in the range $10^{11}\Msun$ to $2\times10^{12}\Msun$.
In this range the mass of the satellites are typical much smaller
than the centrals and major galaxy-galaxy mergers are negligible \citep{LZ14b}. 
For more massive galaxies, Eqs\,\ref{eq:gas:01} and
\,\ref{eq:metal:01} are not sufficient unless the gas and metals 
brought in by mergers are properly taken into account.
Another reason for not to extending to higher halo mass
is the limitation of the observational constraints we use.
The star formation history\,\ref{constraints:sfh}
of more massive galaxies, most of which are quenched, 
is an average over star formation and quenched galaxies,
while the gas phase metallicity\,\ref{constraints:metal}
are limited to star forming galaxies.

\subsection{The Halo Assembly History}
\label{model:hah}

Our empirical model follows galaxy evolution in the context of 
realistic halo assembly histories. The assembly of individual 
dark matter halos is modeled using the halo merger tree 
generator proposed by \citet{Parkinson08}. This is a Monte-Carlo 
model based on a modified treatment of the 
extended Press-Schechter formalism that is calibrated with 
$N$-body simulations \citep[see][]{Cole08}. 
As shown in \citet{Jiang14}, the merger trees
obtained with this method match those obtained with
high-resolution numerical simulations.

Given a halo mass and a redshift, we only follow the average 
assembly history instead of individual merger trees.
The average assembly history is obtained by averaging over 
the main-branch progenitors of different trees.
The intrinsic scatter in both the models and the observational 
constraints, such as the gas phase metallicity-stellar mass 
relation and the gas mass fraction-stellar mass relations,
is ignored for simplicity.

\subsection{The Intrinsic Metal Yield}
\label{model:yield}

\begin{table}
 \centering
 \caption{The oxygen yield as a function of initial stellar metallicity.
          Results obtained from two different stellar evolution models are presented: P98 is
          for \citet{Portinari98} and K06 for \citet{Kobayashi06}.
          Chabrier IMF \citep{Chabrier03} is used in both models.}
 \begin{tabular}{cccc}
 \hline
 \hline
  Model  & $Z_{\rm i}=0.0004$ & $Z_{\rm i}=0.004$ & $Z_{\rm i}=0.02$ \\ 
 \hline
  P98  & 0.0168 & 0.0180 & 0.0163 \\
 \hline 
  K06  & 0.0134 & 0.0110 & 0.0103 \\
 \hline
 \hline 
\end{tabular}
\label{yield}
\end{table}

The intrinsic yield of a simple stellar population can be
estimated from the stellar IMF 
and the adopted stellar evolution model:
\begin{equation}
y = \int_{m_u}^{m_l} m p(m) \phi(m){\rm d}m\,,
\end{equation}
where $\phi(m)$ is the IMF and $p(m)$ is the mass 
fraction of certain metals produced by stars of an initial mass $m$. 
Here we adopt two models,  one is from \citet{Portinari98}
and the other is from \citet{Kobayashi06}.
Table~\ref{yield} lists the yield of oxygen for different initial
metallicities of the stellar population.
For both models the oxygen yield depends mildly on the 
initial stellar metallicity. However, the variance between 
different models is considerable. The yield predicted by 
the \citet{Kobayashi06} model is about $2/3$ of 
that by the \citet{Portinari98} model.
In this paper, we choose the \citet{Portinari98} model as 
our fiducial model, since it is consistent with a broad range
of stellar evolutions models in the literature \citep{Peeples14}.
The consequence of using a smaller 
yield will be discussed whenever needed.


\section{Observational Constraints}
\label{constraints}

\begin{figure*}
 \centering
 \begin{minipage}{0.45\linewidth}
 \includegraphics[width=\linewidth]{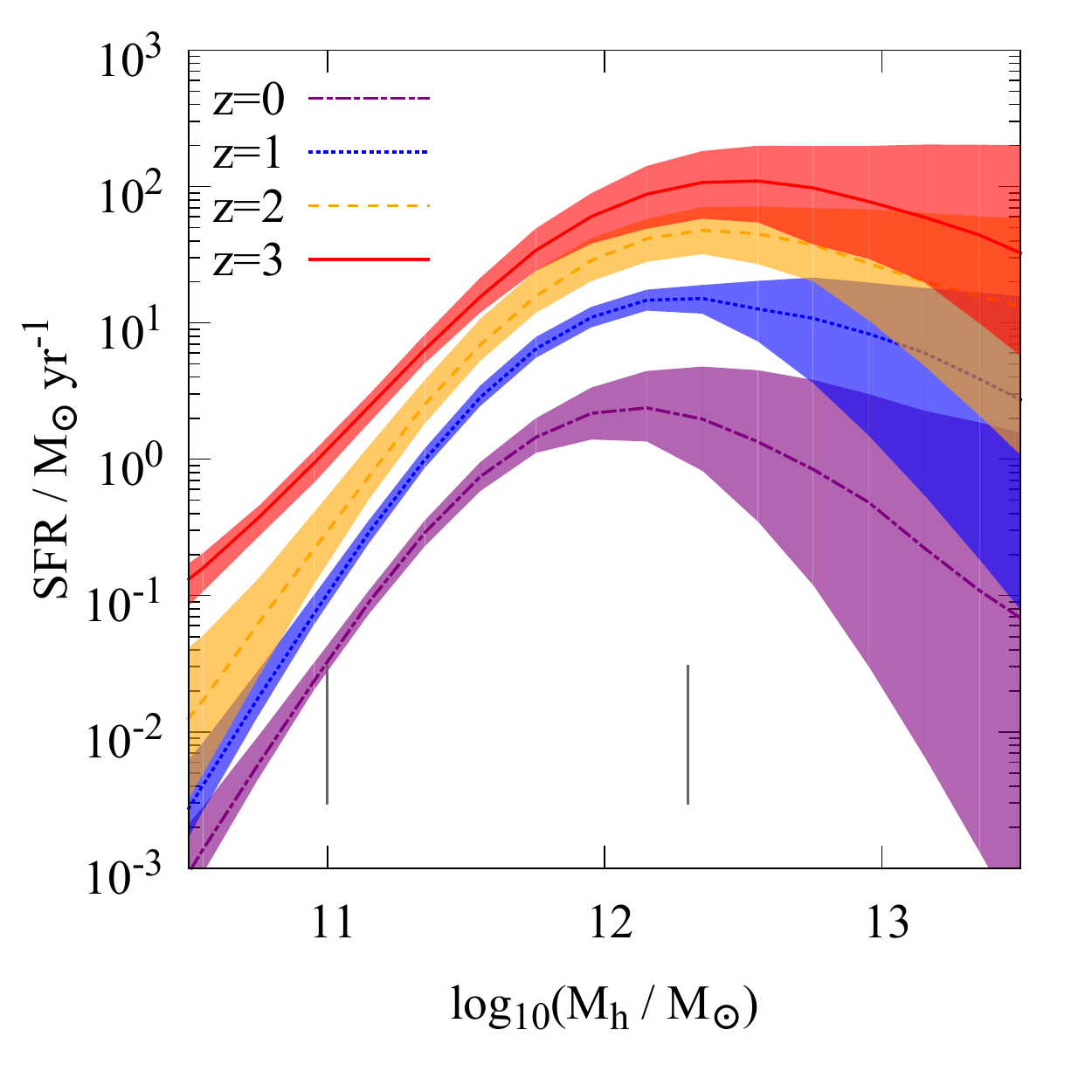}
 \end{minipage}
 \begin{minipage}{0.45\linewidth}
 \includegraphics[width=\linewidth]{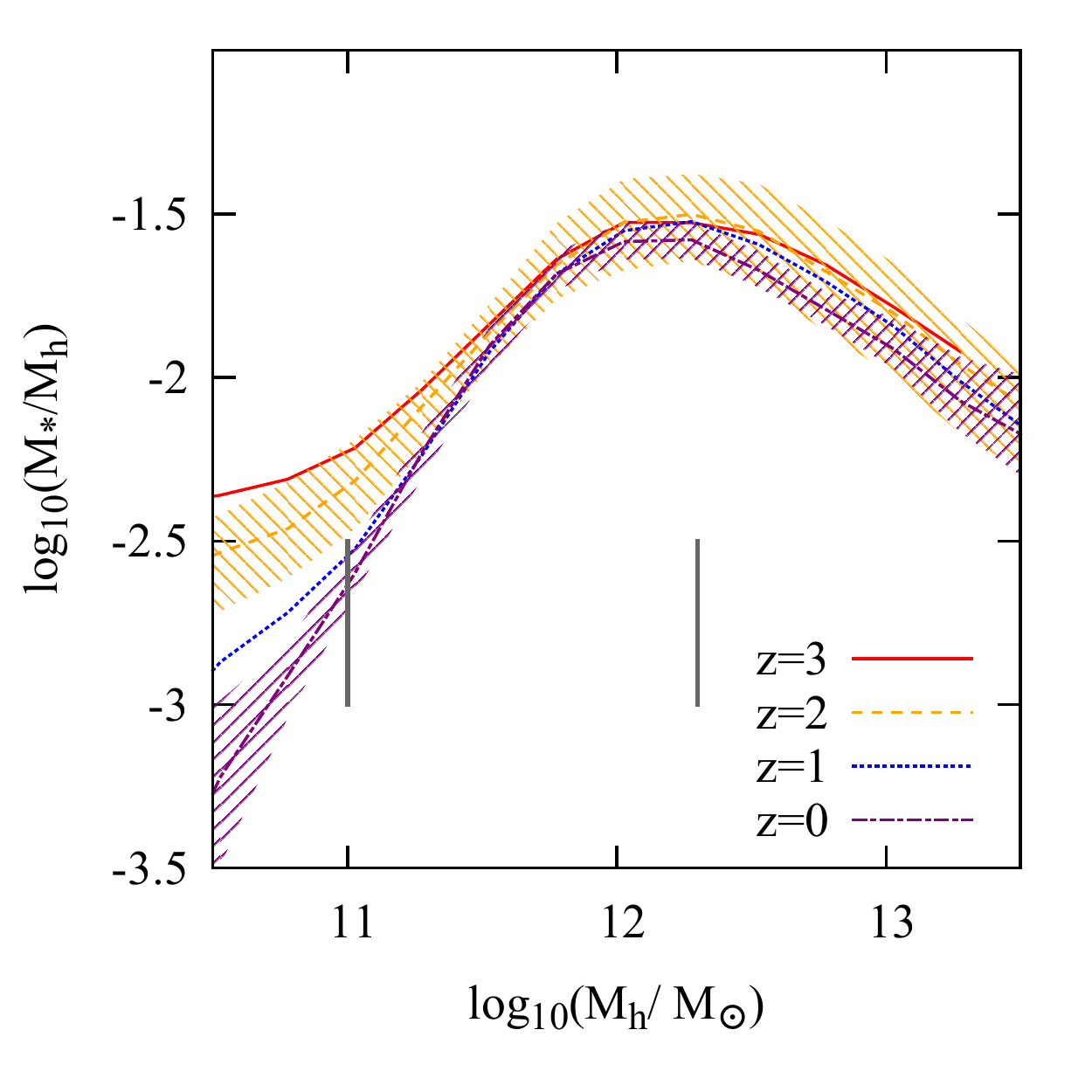}
 \end{minipage}
 \caption{The dependence of SFR (left) and stellar mass to halo mass ratio (right) 
          on halo mass and redshift. The bands represent the $95\%$ credible 
          intervals. The grey vertical lines bracket the halo mass range
          we focus on in this work.}
 \label{sfh}
\end{figure*}

\subsection{Star Formation History}
\label{constraints:sfh}

We make use of the star formation rate (SFR) - halo mass relation obtained by
\citet{LZ14a} and \citet{LZ14b} as one of our constraints.
Specifically, we adopt Model III, in which the star formation rate 
is written as
\begin{equation}
 \label{eqn_general}
 \Psi = {\cal E} {f_{\rm b} M_{\rm h} \over\tau}
        (X+1)^{\alpha}  \left(\frac{X+\mathcal{R}}{X+1}\right)^{\beta} 
        \left(\frac{X}{X+\mathcal{R}}  \right)^{\gamma} 
        \,,
\end{equation}
where ${\cal E}$ is a free parameter that sets the overall efficiency,
$f_{\rm b} = \Omega_{\rm b,0}/\Omega_{\rm m,0}$ is the cosmic baryon
mass fraction, and $\tau = (1/10\,H_0) \, (1+z)^{-3/2}$ roughly
describes the dynamical timescale of halos at redshift $z$.
The quantity $X$ is defined as $X \equiv M_{\rm h} / M_{\rm c}$,
where $M_{\rm c}$ is a characteristic mass, and $\mathcal{R}$ is a
positive number that is smaller than $1$.  Hence, the SFR depends on
halo mass through a piecewise power law, with $\alpha$, $\beta$, and
$\gamma$ being the three power indices in the three different mass
ranges separated by the two characteristic masses, $M_{\rm c}$ and
$\mathcal{R} M_{\rm c}$. In this model, the index $\alpha$ 
is assumed to depend on redshift according to
\begin{equation}\label{eqn_alpha}
  \alpha = \alpha_{0} (1+z)^{\alpha'} \,, 
\end{equation}
and $\gamma$ according to
\begin{equation}\label{eqn_gamma}
  \gamma = 
  \begin{cases}
    \gamma_{\rm a}   \, & \text{if}\, z < z_{c} \\
    (\gamma_{\rm a}-\gamma_{\rm b})
    \left(\frac{z+1}{z_{c}+1}\right)^{\gamma'} + 
     \gamma_{\rm b}  \, & \text{otherwise}\,.
  \end{cases}
\end{equation}
Thus $\gamma$ changes from $\gamma_{\rm b}$ 
at high-$z$ to $\gamma_{\rm a}$ at low-$z$, with 
a transition redshift $z_{\rm c}$.
  
The model is constrained using the galaxy stellar mass function (SMF)
at $z\approx 0$ from \citet{Baldry12}, the SMFs at $z$ between $1$ and $4$ 
from \citet{Santini12} and the z-band cluster galaxies luminosity function
by \citet{Popesso06}. The constrained parameters can be found in
\citet{LZ14b}. The SFR as a function of halo mass and 
redshift and the consequent stellar mass to halo mass ratio are shown
in Figure~\ref{sfh} for reference, with the bands representing 
the inferential uncertainty. In the mass range we are interested in
here (between the two vertical grey lines),
this uncertainty is quite small. We therefore ignore the scatter
and use the best fit parameters to characterize the star 
formation histories.

\subsection{Gas Phase Metallicity}
\label{constraints:metal}

\begin{figure}
 \centering
 \includegraphics[width=0.9\linewidth]{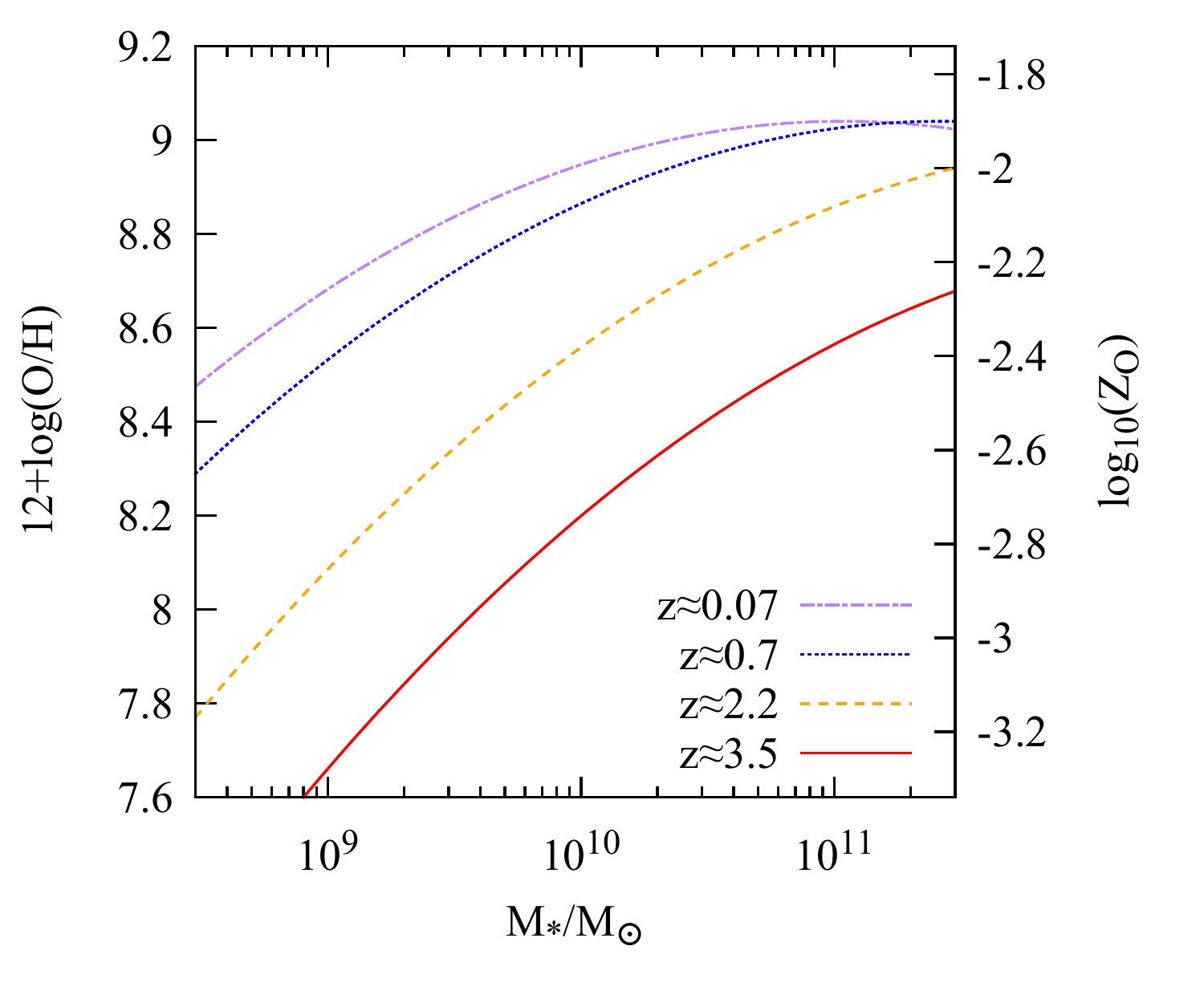}
 \caption{Gas phase metallicity compiled by \citet{Maiolino08}. 
          The $y$-axis on the right is the corresponding oxygen mass fraction.}
 \label{metal}
\end{figure}

Another observational constraint adopted in this paper is the gas phase metallicity,
which is usually measured from the emission lines of the HII regions
of star forming galaxies. 
In this paper, we adopt the metallicity measurements
compiled by \citet{Maiolino08}. 
Figure~\ref{metal} shows the metallicity-stellar 
mass relations obtained from their fitting formula.
It is important to realize that the metallicity measurements 
have significant systematic error. 
For local galaxies the random error in the measurements
is only about 0.03 dex \citep{Tremonti04}, 
but the systematic uncertainty due to different ways to convert 
the emission lines into abundances is as large as 0.7 dex \citep{Kewley08}. 
There are two ways to estimate the metal abundance from 
such observations: the electron temperature ($T_{\rm e}$) 
method and the theoretical method. 
In the $T_e$ method, the ratio between the 
${\rm [OIII]}\lambda4363$ auroral line and 
${\rm [OIII]}\lambda5007$ is used to estimate the mean 
electron temperature, which is in turn used to estimate
the oxygen abundance \citep{Peimbert69}. 
In the theoretical method, a sophisticated photoionization model 
is fit to the strong line ratios, such as
$R_{\rm 23} = ({\rm [OII]}\lambda3737+{\rm
  [OIII]}\lambda4959,5007)/{\rm H\beta}$.
Empirical calibrations based on the two methods often 
show a discrepancy as large as $0.7$ dex. 
 \citet{Stasinska05} pointed out that 
due to the temperature fluctuation or gradient in 
high metallicity [$12+\log_{10}({\rm O/H})>8.6$] HII regions,
the $T_{e}$ method can underestimate the metallicity by 
as much as $0.4$ dex. Meanwhile the systematics in the 
photoionization modeling can be as large as 0.2 dex \citep{Kewley08}.
In \citet{Maiolino08}, both of the two methods described above
are used to derive the relations between the strong line ratios and metallicity.
Specifically, the $T_{e}$ method is only applied to metal poor galaxies
($12+\log_{10}({\rm O/H})<8.6$) to avoid bias. 
The empirical calibrations derived in this way
cover a large metallicity range and therefore can be 
applied to galaxies over a large redshift range.

\subsection{Gas Fraction in Local Galaxies}
\label{constraints:gas}

\begin{figure*}
 \centering
 \begin{minipage}{0.45\linewidth}
 \includegraphics[width=\linewidth]{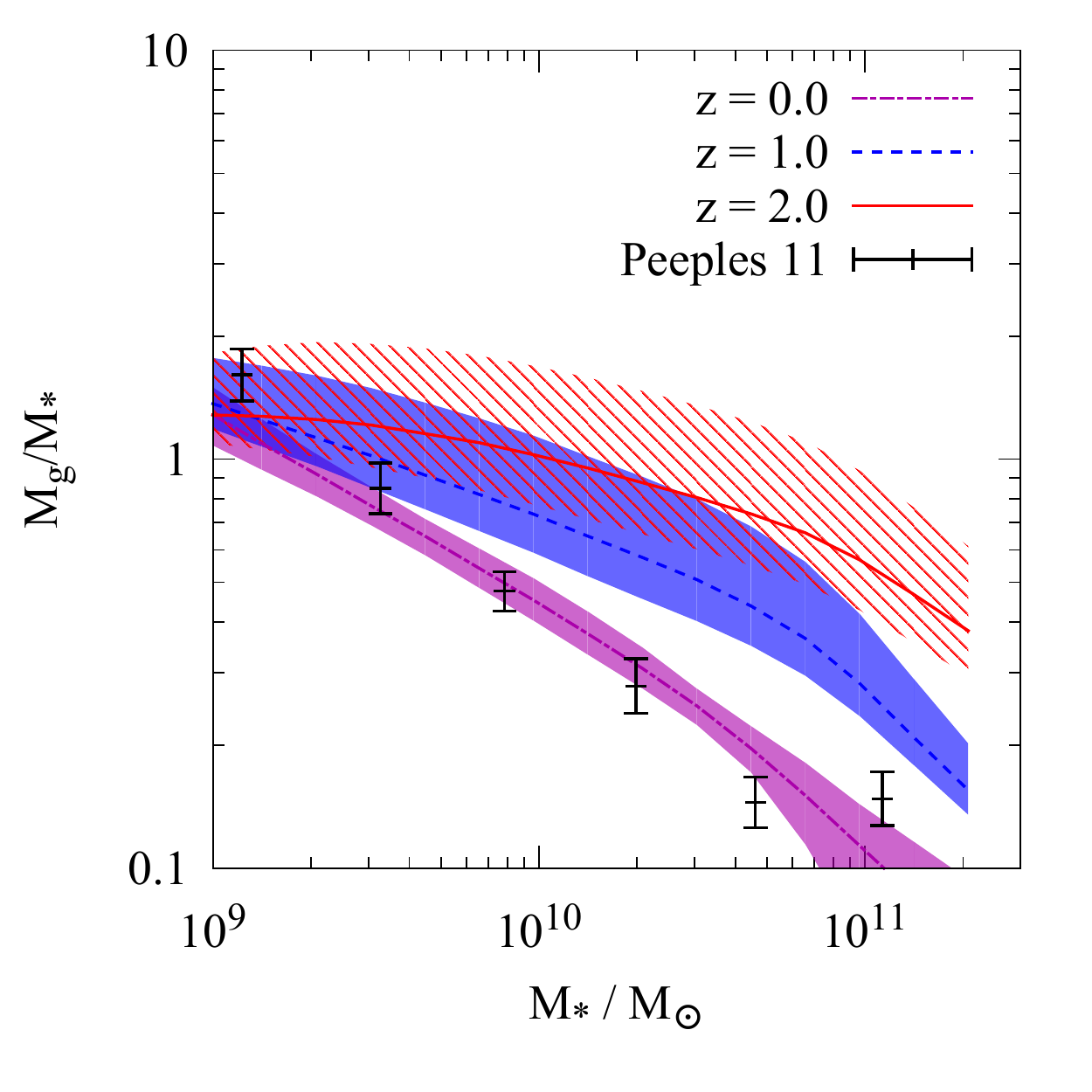}
 \end{minipage}
 \begin{minipage}{0.45\linewidth}
 \includegraphics[width=\linewidth]{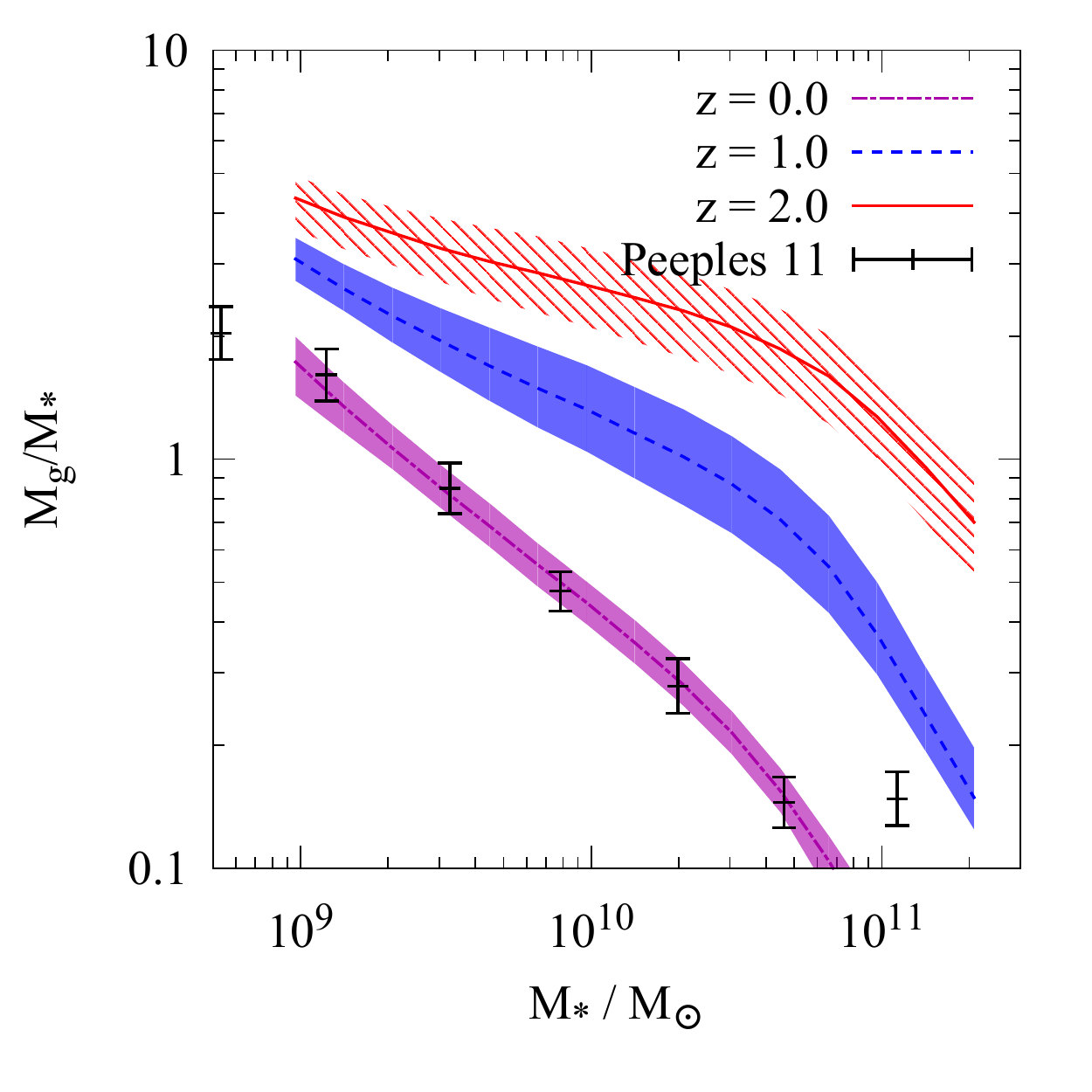}
 \end{minipage}
 \caption{Cold gas to stellar mass ratio as a function of stellar mass
    at different redshifts calculated using the Kennicutt-Schmidt Law (left) 
    and the Krumholz model (right).
    The lines are the predictions of the best fitting model in
    Table\,\ref{SFL_param} and the bands are obtained by marginalizing
    the uncertainties in the parameters.
    The data points are compilation of \citet{Peeples11} from different observations of local
    galaxies.} 
 \label{Fg}
\end{figure*}

In addition, we also include the observations
of gas contents in local galaxies compiled by \citet{Peeples11} 
as a constraint.  
The data points in Figure~\ref{Fg}, which show the total gas mass
to stellar mass ratios, are taken from \citet{Peeples11}. 
The binned data points are compiled from several different sources, 
taken into account HI, helium and molecular hydrogen. 
Here both the mean relation and the uncertainties, taken as random errors,  
are used in the data constraint.


\section{Evolution of cold gas content of galaxies}
\label{gas_content}

Given the observational constraints for 
the star formation histories in \S\ref{constraints:sfh}
and for the gas phase metallicity in \S\ref{constraints:metal},
we can solve Eq.~(\ref{eq:star}) to obtain the gas mass $M_{\rm g}$
by adopting specific models for the star formation rate and 
for the structure of the cold gas distribution.
In this section, we first introduce the star formation
(\S\ref{gas_content:sfl}) and disk structure 
(\S\ref{gas_content:disk}) models we adopt, 
we then show the predictions for the cold gas mass in high redshift
galaxies (\S\ref{gas_content:mass}).

\begin{table}
 \centering
 \caption{Parameters in the Kennicutt-Schmidt model,
 $A_{\rm K}$ and $\Sigma_{\rm c}$, and in the Krumholz model,
 $\tau_{\rm sf}$ and $c$, tuned together with the disk-size
 parameter, ${\cal L}$,  to match the gas mass to stellar mass ratio of 
 local galaxies (data points in Figure\,\ref{Fg}).
The second column lists the fitting results and 
the third column lists the equations which define the
corresponding parameters.}
  \begin{tabular}{ccc}
 \hline
 \hline
  parameter & value & \\ 
 \hline
  $A_{\rm K}$                      & $(2.5\pm1.2)\times10^{-4}$ & \multirow{4}{*}{Eqs.\,(\ref{kennicutt_a}) 
  - (\ref{kennicutt_b})}\\  
  $/{\rm M_{\odot}yr^{-1}pc^{-2}}$ &                    & \\ [1.0ex] 
  $\Sigma_{\rm c}$                 & $9.8\pm2.6$        & \\
  $/{\rm M_{\odot}pc^{-2}}$        &                    & \\ [1.0ex]
  $\mathcal{L}$                    & $2.1\pm0.4$        & Eq\,(\ref{disk_size}) \\ [1.0ex]                        
 \hline
  $\tau_{\rm sf}/{\rm Gyr}$ & $2.5\pm0.3$ & \multirow{2}{*}{Eqs.\,(\ref{krumholz_a}) - (\ref{krumholz_c})} \\ [1.0ex]
  $c$                       & $2.6\pm1.5$ & \\ [1.0ex]
  $\mathcal{L}$             & $3.3\pm1.0$ & Eq.\,(\ref{disk_size}) \\
 \hline
 \hline 
\end{tabular}
\label{SFL_param}
\end{table}

 \begin{figure*}
  \centering
  \begin{minipage}{0.45\linewidth}
  \includegraphics[width=\linewidth]{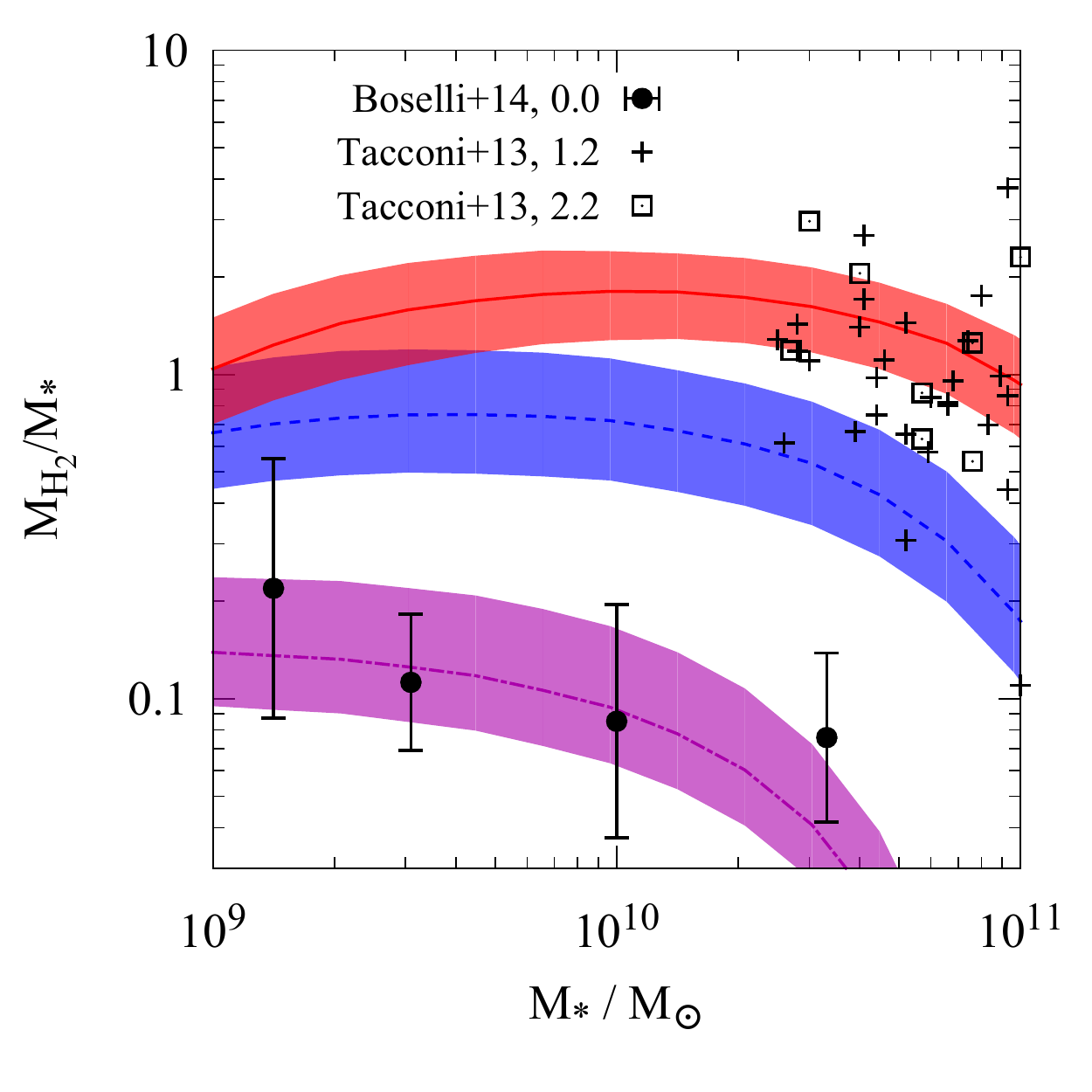}
  \end{minipage}
  \begin{minipage}{0.45\linewidth}
  \includegraphics[width=\linewidth]{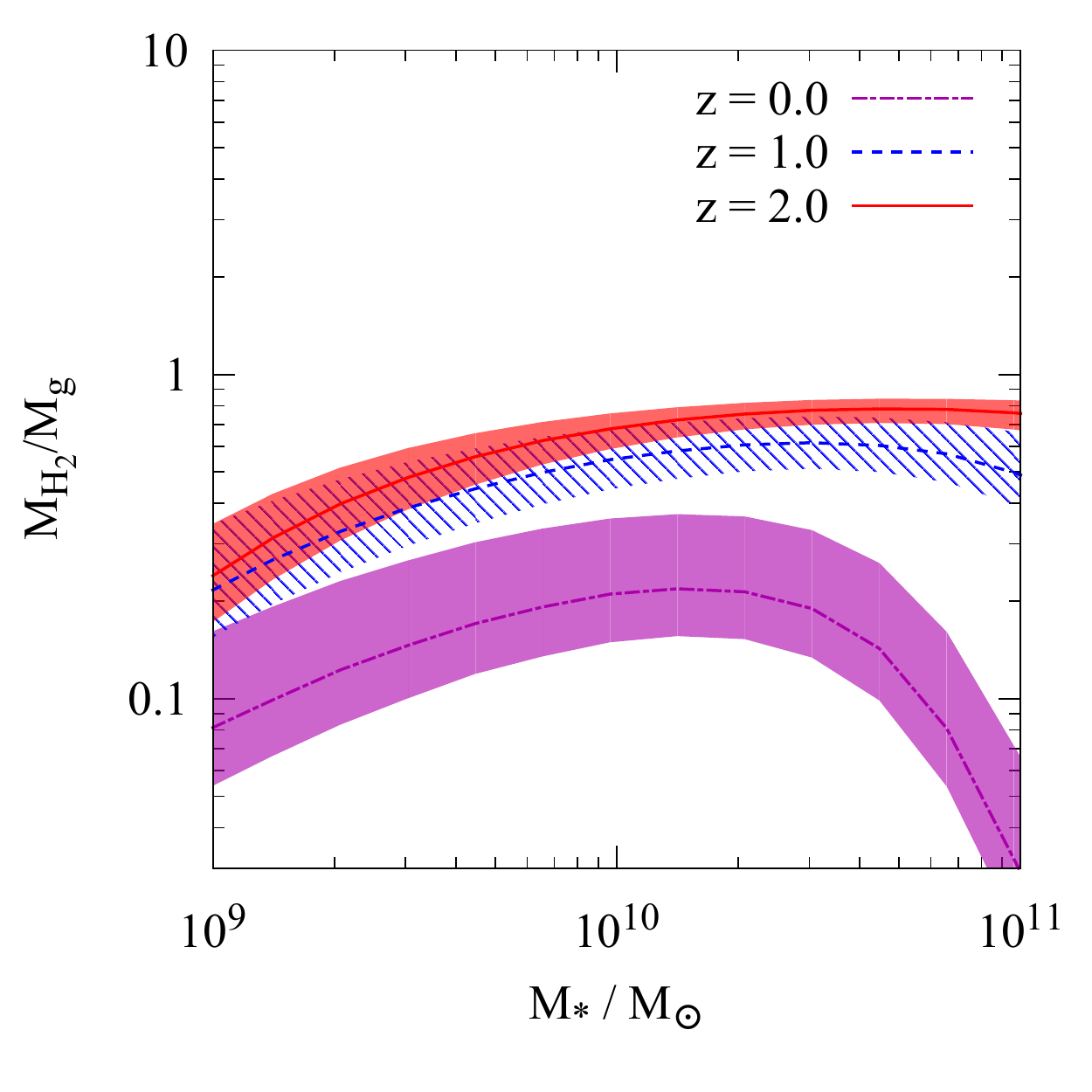}
  \end{minipage}
  \caption{
           Molecular gas to stellar mass ratio (left), molecular gas to total gas mass ratio
           (right) as a function of stellar mass, all predicted by the Krumholz model.
           The lines are the predictions of the best fitting model in
           Table\,\ref{SFL_param} and the bands are obtained by marginalizing
           the uncertainties in the parameters.
           The data points from \citet{Tacconi13} are individual galaxies 
           the binned data points for local galaxies are from \citet{Boselli14}.}
  \label{H2}
 \end{figure*}
   
 \begin{figure*}
  \centering
  \begin{minipage}{0.48\linewidth}
  \includegraphics[width=\linewidth]{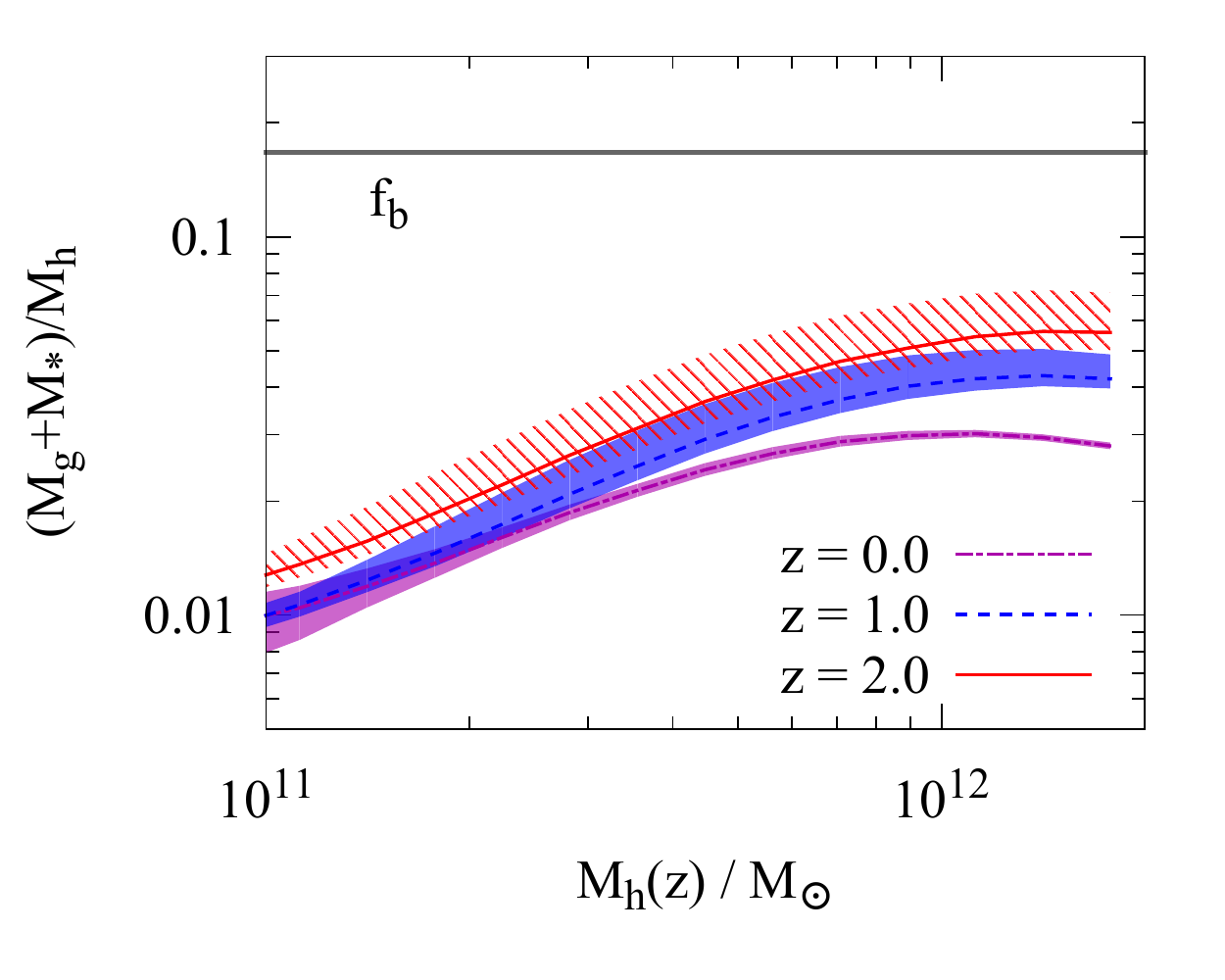}
  \end{minipage}
  \begin{minipage}{0.48\linewidth}
  \includegraphics[width=\linewidth]{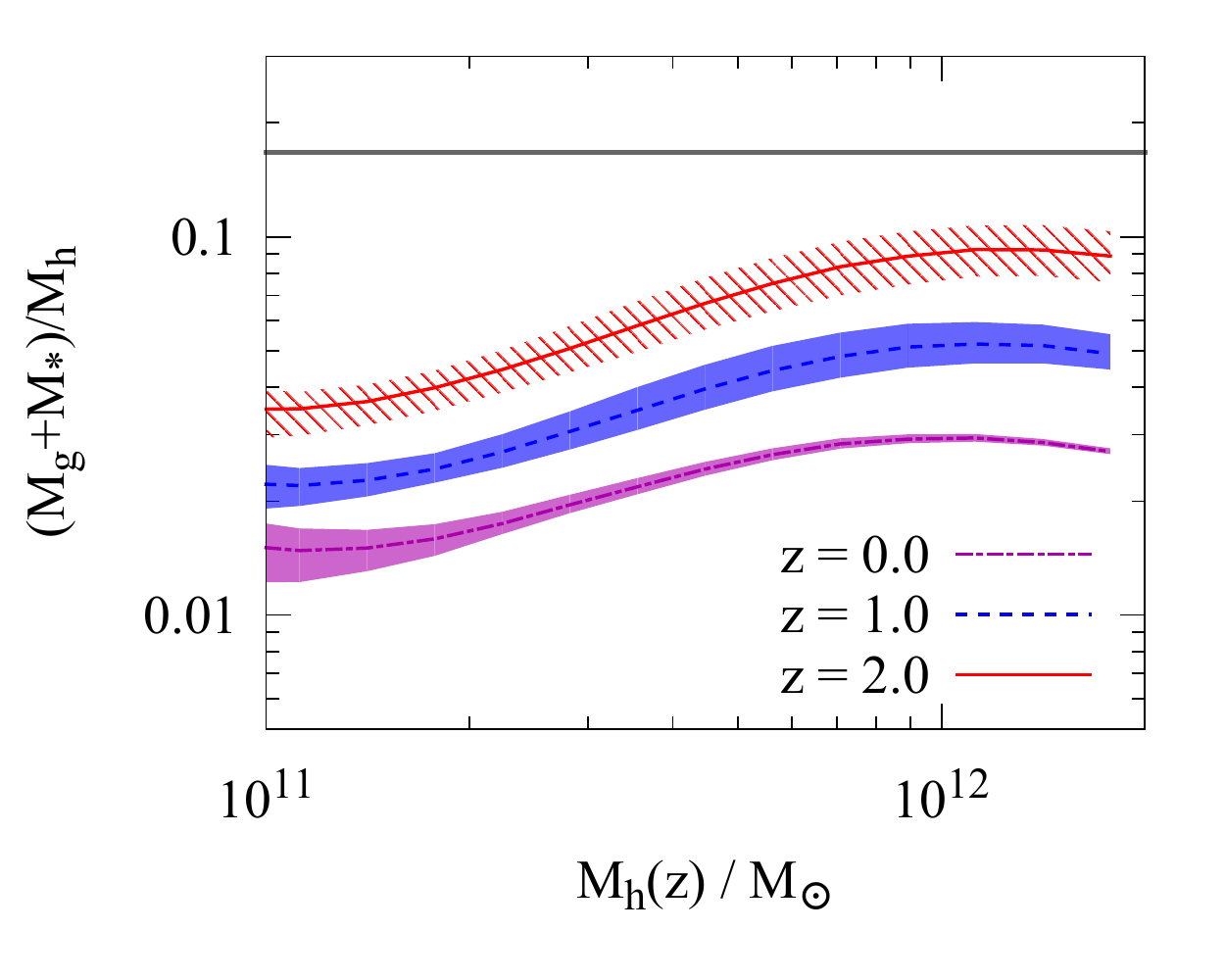}
  \end{minipage}
  \caption{Baryon mass (in both stars and cold gas)
           to halo mass ratio as a function of halo mass
           calculated using the Kennicutt-Schmidt Law (left) and the Krumholz Law (right).
           The horizontal lines indicate the universal baryon fraction.
           The lines are the predictions of the best fitting model in
           Table\,\ref{SFL_param} and the bands are obtained by marginalizing
           the uncertainties in the parameters.} 
 \label{Mb}
 \end{figure*}

\subsection{The Star Formation Models}
\label{gas_content:sfl}
 
We consider two different star formation models widely adopted
in the literature. The first is the Kennicutt-Schmidt 
Law \citep{Kennicutt98}, an empirical relation between
the SFR surface density, $\Sigma_{\rm SFR}$, 
and the cold gas surface density, $\Sigma_{\rm g}$,
\begin{equation}
\label{kennicutt_a}
\Sigma_{\rm SFR} = 
A_{\rm K} \left(\frac{\Sigma_{\rm g}}{\rm M_{\odot}pc^{-2}}\right)^{N_{\rm K}}\,,
\end{equation}
where the power index $N_{\rm K} \approx 1.4$, and 
$A_{\rm K}$ is a constant amplitude.
In this model, star formation is assumed to occur 
only in cold gas disks where the surface density 
exceeds a threshold $\Sigma_{\rm c}$. 
Assuming the cold gas disk follows an exponential profile, 
the total SFR can be obtained as, 
\begin{equation}
\label{kennicutt_b}
 \Psi =
 \begin{cases}
   \frac{2\pi A_{\rm K}\Sigma^{N_{\rm K}} R_{\rm g}^{2}}{N_{\rm K}^{2}} 
   \left[ 1-\left(1+N_{\rm K}\frac{r_{\rm c}}{R_{\rm g}} \right) 
          \exp\left(-N_{\rm K}\frac{r_{\rm c}}{R_{\rm g}}\right) 
   \right] \\
   ~~~~~~~~ \text{if} ~ \Sigma_{0}\ge \Sigma_{\rm c}            \\
   0       \\
   ~~~~~~~~ \text{if} ~ \Sigma_{0} <  \Sigma_{\rm c}            \, ,
 \end{cases}
\end{equation}
where $R_{\rm g}$ is the scale radius of the disk, 
$\Sigma_{0} \equiv \frac{M_{\rm g}}{2\pi R_{\rm g}^{2}}$ 
is the surface density at the disk center, and
$r_{\rm c} = \ln\left(\Sigma_{0}/\Sigma_{\rm c}\right)R_{\rm g}$
is the critical radius, within which star formation can happen.
Both $A_{\rm K}$ and $\Sigma_{\rm c}$ are treated as 
free parameters to be determined by observational constraints.

The other star formation model adopted here is the one 
proposed by \citet{Krumholz08, Krumholz09}, in which
the SFR is assumed to be directly related to the properties of 
the molecular cloud: 
\begin{equation}
\label{krumholz_a}
 \Sigma_{\rm SFR} = \frac{\epsilon_{\rm ff}}{t_{\rm ff}}\Sigma_{\rm H_2}\,,
\end{equation}
where $\Sigma_{\rm H_2}$ is the surface density of molecular hydrogen, 
and  $t_{\rm ff}$ is the local free fall time scale. 
The ratio $\epsilon_{\rm ff}/t_{\rm ff}$
depends on the total gas surface density:
\begin{equation}
\label{krumholz_b}
 \frac{\epsilon_{\rm ff}}{t_{\rm ff}} = \frac{1}{\tau_{\rm sf}}
 \begin{cases}
  \left(\Sigma_{\rm g}/85{\rm M_{\odot} pc^{-2}}\right)^{-0.33} \, & \text{if} ~ \Sigma_{\rm g} <   85{\rm M_{\odot}pc^{-2}} \\
  \left(\Sigma_{\rm g}/85{\rm M_{\odot} pc^{-2}}\right)^{ 0.33} \, & \text{if} ~ \Sigma_{\rm g} \ge 85{\rm M_{\odot}pc^{-2}} \,,
 \end{cases}
\end{equation}
where $\tau_{\rm sf}$ is a constant, treated as a free parameter.
The fraction of molecular gas, $f_{\rm H_2}=\Sigma_{\rm H_2}/\Sigma_{\rm g}$, 
depends primarily on the surface density and metallicity of the cold gas, and is modeled as
\begin{eqnarray}
\label{krumholz_c}
\label{eq:H2frac}
 f_{\rm H_2} = 
 \begin{cases}
  1 - \frac{3}{4}\left(\frac{s}{1+0.25s}\right) \, & \text{if} ~ s \le 2 \\
  0 \, & \text{if} ~ s > 2 \,
 \end{cases} \\ 
 s = \frac{\ln(1+0.6\chi+0.01\chi^{2})}{0.6\tau_{\rm c}} \nonumber \\
 \chi = 3.1\frac{1+Z_{o}^{0.365}}{4.1} \nonumber \\
 \tau_{\rm c} = 320cZ_{o}\frac{\Sigma_{\rm g}}{\rm g/cm^{2}} \nonumber \, .
\end{eqnarray}
Here $Z_{o}$ is the metallicity normalized to the solar value,
$c$ is a constant treated as a free parameter, and $s=2$ 
defines a threshold surface density 
for the formation of molecular hydrogen. Note that $s$ is roughly
inversely proportional to the gas phase metallicity, 
so that a high metallicity corresponds to a lower surface density 
threshold.

\subsection{Disk Size}
\label{gas_content:disk}

To determine the distribution of cold gas, 
we assume that the cold gas disk follows an exponential radial profile 
with a disk size proportional to the stellar disk. 
We estimate the stellar disk size using the empirical 
size-stellar mass relation obtained by \citet{Dutton11}
for nearby galaxies ($z\approx0.1$), 
\begin{equation}
 R_{\rm 50} = R_{\rm 0} \left(\frac{M_{\star}}{M_{\rm 0}}\right)^{\alpha}
     \left[\frac{1}{2}+\frac{1}{2}\left(\frac{M_{\star}}
     {M_{\rm 0}}\right)^{\gamma}\right]^{(\beta-\alpha)/\gamma}\,,
\end{equation}
where $R_{50}$ is the half light radius of the stellar disk, 
$\log_{10}(M_{\rm 0}/M_{\odot}) = 10.44$, $\log_{10}(R_{\rm 0}/{\rm kpc}) = 0.72$,
$\alpha = 0.18$, $\beta = 0.52$ and $\gamma=1.8$.
With the assumption that the shape of the relation holds at all redshifts,
the time evolution of the disk size is given by the offset
\begin{equation}
 \Delta \log_{10}(R_{\rm 50}) = 0.018 - 0.44\log_{10}(1+z).
\end{equation}
This redshift dependence is slightly shallower than the more
recent observational calibration by \citet{vdWel14}, 
which is $\propto (1+z)^{-0.75}$, but our results are not sensitive to it.
As shown by \citet{Dutton11} the star formation activity 
typically has a more extended distribution than the stellar disk,
with a size about two times the stellar disk, and the relation 
does not evolve strongly with time. The gas disk is traced by 
the star formation to some extent. 
Using a sample of local galaxies that covers a broad range of 
stellar mass and morphological types, \citet{Kravtsov13} showed
that the sizes of the cold gas disks are typically larger 
than the stellar disks by a factor of $\approx2.6$.
Investigating a semi-analytic model that implements 
detailed treatments of gas distribution 
and star formation, \citet{LY14} found that the size ratio between 
the cold gas disk and the stellar disk ranges from $2$ to $3$ for 
galaxies with mass in the range considered here. 
In our model we therefore assume that 
\begin{equation}
 \label{disk_size}
 R_{\rm g} = {\cal L} R_{\star}\,,
\end{equation} 
with ${\cal L}$ treated as a free parameter to be tuned along with some 
other parameters in the star formation models (Table\,\ref{SFL_param})
to match the observed gas fraction of local galaxies.

\subsection{The Cold Gas Contents}
\label{gas_content:mass}

To make use of the models described above, 
we first calibrate the parameters in the star formation laws
and the gas disk size parameter ${\cal L}$ 
using the observed gas mass/stellar mass ratio
of local galaxies \citep{Peeples11}. The best fits and 
the $1~\sigma$ uncertainties of the tuned parameters 
are listed in Table~\ref{SFL_param}. 
The predicted cold gas contents as functions of stellar mass 
are shown in Figs.\,\ref{Fg} and \ref{H2}.

Both of the star formation laws can successfully reproduce 
the cold gas fraction of local galaxies by tuning the corresponding 
model parameters.
This is in contrast with the finding of \citet{Peeples11} 
that the Schmidt-Kennicut law fails to match the high
gas mass fraction in dwarf galaxies. We find that the critical surface 
density $\Sigma_{\rm c}$, which was not taken 
into account in \citet{Peeples11}, is crucial in reproducing the steep 
gas mass fraction-stellar mass relation.
Similarly, the Krumholz star formation model also has
a critical surface density for molecule formation,
which is roughly inversely proportional to the gas phase 
metallicity [Eq.\,(\ref{eq:H2frac})]. The key difference between 
the two star formation models is that the critical surface 
density in the Schmidt-Kennicutt law is a constant,
while that in the Krumholz model changes with time and 
the mass of the host galaxies.
According to the observed gas phase metallicity (Figure\,\ref{metal}),
the critical surface density in Krumholz model 
increases with redshift. To sustain the same 
amount of star formation, the gas fraction derived 
from this model is thus higher than that 
derived from the Schmidt-Kennicutt law, 
especially for dwarf galaxies with stellar masses $<10^{9}\Msun$.
Using a molecule-regulated star formation model,
\citet{Dutton10} inferred that the gas to stellar mass ratio
changes only weakly with time, in contrast to our results shown in 
Figure\,\ref{Fg}. The major reason for the difference is that
in their model the formation of molecular hydrogen
is determined by the total gas surface density,
while in the Krumholz model the evolution of metallicity plays a crucial role.
Clearly, the gas mass in high redshift
galaxies is sensitive to the assumed star formation model, 
and more models need to be explored and checked
with future observations \citep{Popping12, Popping14}.

The Krumholz model also allows us to infer the gas fraction in molecular phase. 
The left panel in Figure\,\ref{H2} shows the molecular gas
to stellar mass ratio. At $z=0$ the ratio is about $0.1$, 
and it increases by an order of magnitude at $z=2$.
The predictions are consistent with the recent measurements 
from \citet{Boselli14} and \citet{Tacconi13}.
The right panel shows the molecular gas to total gas mass ratio
as a function of stellar mass. 
At $z>1$, most of the gas is in the  molecular phase.

Regardless which star formation model is adopted, the ratio 
between the total baryon mass 
settled in the galaxies and the host halo mass is always much less than 
the universal baryon mass fraction (see Figure\,\ref{Mb}).
This deficit of baryon mass strongly indicates that 
star formation models alone cannot account for the low star
formation efficiency in low-mass halos.  Processes that control 
the gas exchange between the surrounding medium and galactic 
medium in forms of gas inflow and outflow must have played a major
role. In the following section, we infer limits on the 
inflow and outflow rates in low-mass galaxies from 
the constrained star formation histories, cold gas fractions, 
and metallicity measurements. 


\section{Inflow and Outflow}
\label{io}

\begin{figure*}
 \centering
 \includegraphics[width=0.8\linewidth]{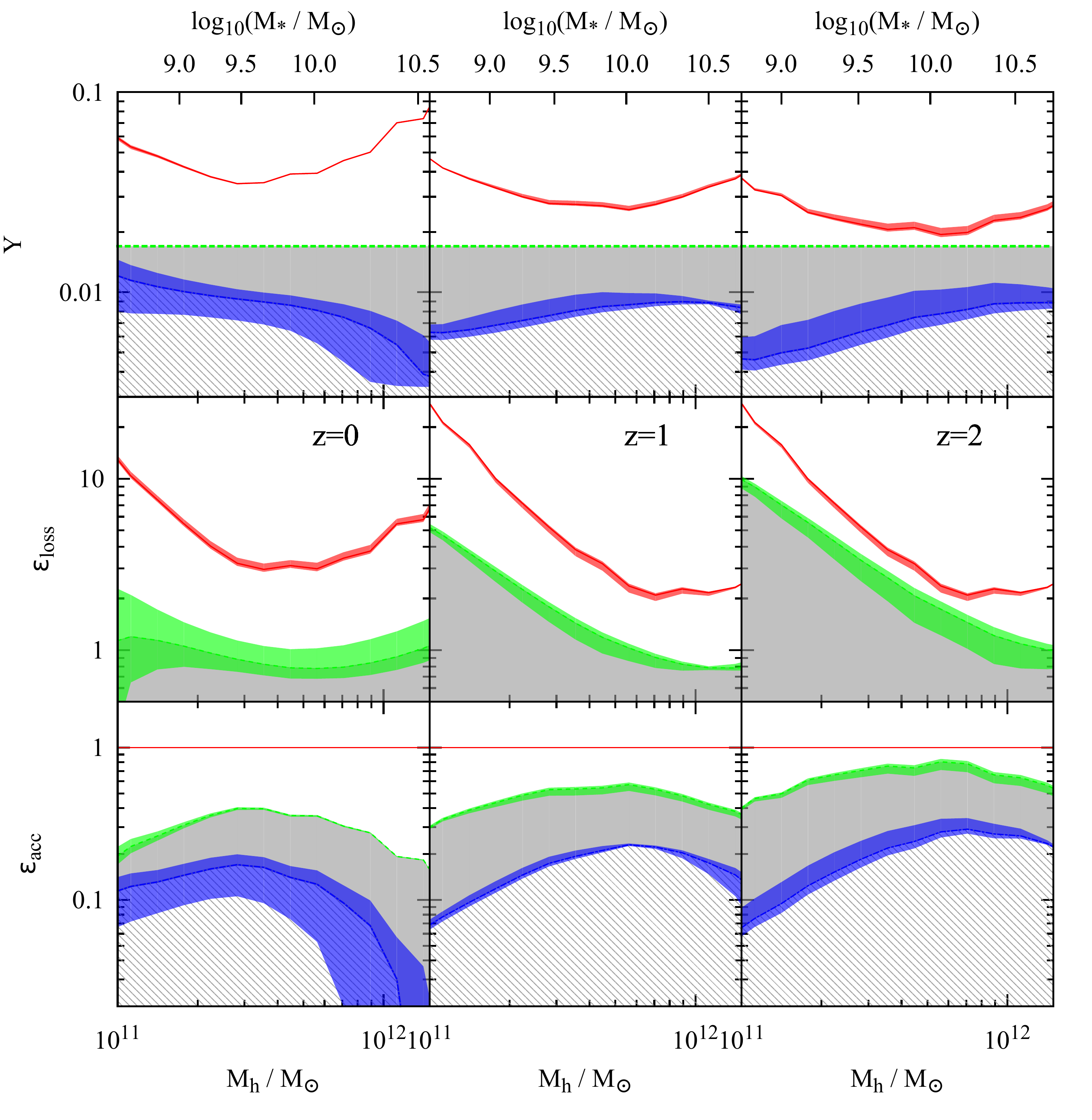}
 \caption{The red lines (bands) are obtained by assuming $\epsilon_{\rm acc} = 1$,
          i.e. galaxies accrete at the maximum rate.
          The green lines (bands) are obtained by assuming $\mathcal{Y} = y$,
          which means full mixing of newly produced metals in the ISM,
          and no recycling or instanteneous recycling of the ejected material.
          The blue lines (bands) are obtained by setting $\epsilon_{\rm loss} = 0$.
          The areas that are not shaded are forbidden by the observational constraints.
          Here the Kennicutt-Schmidt law is assumed. 
          The lines are the predictions of the best fitting model in
          Table\,\ref{SFL_param} and the bands are obtained by marginalizing
          the uncertainties in the parameters. The hatched regions correspond to          
          constraints when $\epsilon_{\rm loss}$ is allowed to be negative (see text).}
 \label{flow_KSv1}
\end{figure*}

\begin{figure*}
 \centering
 \includegraphics[width=0.8\linewidth]{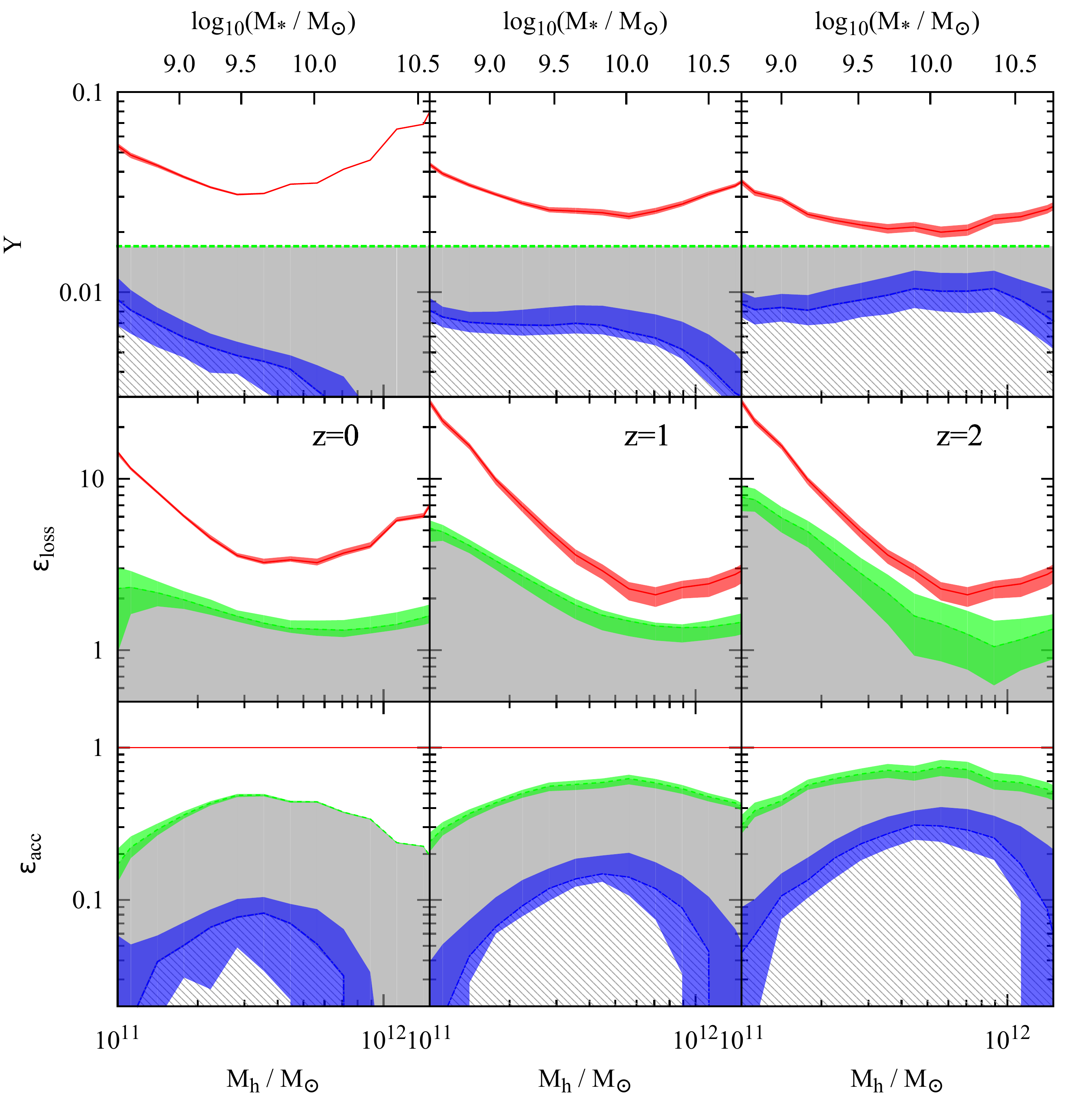}
 \caption{The same as Figure~\ref{flow_KSv1} but here the Krumholz star formation
          model is assumed.}
 \label{flow_KMv1}
\end{figure*}

As described above, the main components of galaxies, such as
halo mass $M_{\rm h}$, stellar mass $M_\star$,
gas mass $M_{\rm g}$ and mass in gas phase metals 
$M_{\rm Z}\equiv M_{\rm g}Z$, and their time derivatives 
can either be obtained directly from observational 
constraints (\S\ref{constraints}) or from modeling (\S\ref{gas_content}).
In this section, we go a step forward by constraining
the terms pertaining to inflow and outflow in 
Eqs.\,(\ref{eq:gas:01}) and (\ref{eq:metal:01}). As we will 
see below, these terms cannot be completely determined, 
but stringent limits can be obtained for them.

To proceed we rewrite Eqs.\,(\ref{eq:gas:01}) and (\ref{eq:metal:01})
in more transparent forms. Since the metallicity of the IGM is 
expected to be much lower than that of the ISM, we set $Z_{\rm IGM}=0$ 
for simplicity. The gas and chemical evolution equations are then reduced to
\begin{eqnarray}
 \label{eq:gas:02}
 \frac{{\rm d} M_{\rm g}}{{\rm d}t} & = &   
     \epsilon_{\rm acc} f_{\rm b}\dot{M_{\rm h}} - \epsilon_{\rm loss}\Psi - (1-R)\Psi\,; \\
 \label{eq:metal:02}
 \frac{{\rm d} M_{\rm Z}}{{\rm d}t} & = & 
    - \epsilon_{\rm loss,Z}\Psi - (1-R)\Psi Z + y\Psi\,,
\end{eqnarray}
where 
\begin{eqnarray}
 \epsilon_{\rm loss} &\equiv& \epsilon_{\rm w} - \epsilon_{\rm r} \, \nonumber\\
	             &\equiv& \frac{\dot{M}_{\rm w}}{\Psi} - \frac{\dot{M}_{\rm r}}{\Psi}
\end{eqnarray}
is the loading factor of net mass loss, and
\begin{equation}
\epsilon_{{\rm loss},Z} \equiv \epsilon_{\rm w}Z_{\rm w} 
- \epsilon_{\rm r}Z_{\rm r}
\end{equation}
is the loading factor of net metal loss.
With some combinations and re-arrangements, 
Eqs.\,(\ref{eq:gas:02}) and (\ref{eq:metal:02}) can be written as
\begin{eqnarray}
 \label{eq:gas:03}
 1-R+\epsilon_{\rm loss} & = & \epsilon_{\rm acc} \mathcal{E}_{\rm SF}^{-1} 
     \left(1 - \epsilon_{\rm acc}^{-1}\frac{\dot{M}_{\rm g}}{f_{\rm b}\dot{M}_{\rm h}}\right)\,; \\
 \label{eq:metal:03}
 \frac{\mathcal{Y}}{Z}   & = & \epsilon_{\rm acc} \mathcal{E}_{\rm SF}^{-1} 
     \left(1 + \epsilon_{\rm acc}^{-1}  \frac{M_{\rm g}}{f_{\rm b}\dot{M}_{\rm h}} \frac{\dot{Z}}{Z}\right) \,,
\end{eqnarray}
where $\mathcal{E}_{\rm SF} \equiv \Psi/(f_{\rm b}\dot{M}_{\rm h})$ 
is the star formation efficiency, which is constrained with the
empirical model of \citet{LZ14a}, and  
\begin{equation}
\mathcal{Y} \equiv y - \epsilon_{\rm w}\left(Z_{\rm w}-Z\right) 
+ \epsilon_{\rm r} \left(Z_{\rm r}-Z\right)\,.
\end{equation}
This quantity can be interpreted as the ``net yield".
For instance, the second term $\epsilon_{\rm w}\left(Z_{\rm w}-Z\right)$
represents the metals taken away by the galactic
wind without being mixed with the ISM.  

The above equations are general and are used to make model predictions 
to be described below. Before presenting the results, let us look at these 
equations under certain approximations, which will help us to understand
the results obtained from the full model and to make connections to results 
obtained earlier under similar approximations.  
Since $M_{\rm g}$ is typically much smaller than $f_{\rm b}M_{\rm h}$,
as shown in Figure~\ref{Mb}, $\dot{M}_{\rm g}/(f_{\rm b}\dot{M}_{\rm h})$
in Eq.\,(\ref{eq:gas:03}) and $[M_{\rm g}/(f_{\rm b}\dot{M}_{\rm h})]({\dot Z}/Z)$
in Eq.\,(\ref{eq:metal:03}) are expected to be much less than unity.
Thus, if $\epsilon_{\rm acc}$ is of the order of unity or 
$\epsilon_{\rm acc} \gg [M_{\rm g}/(f_{\rm b}\dot{M}_{\rm h})]({\dot{Z}}/Z)$ and
$\epsilon_{\rm acc} \gg \dot{M_{\rm g}}/(f_{\rm b}\dot{M}_{\rm h})$, 
the above equations can be simplified to
\begin{eqnarray}
 \label{eq:gas:approx}
 1-R+\epsilon_{\rm loss} & \approx & \epsilon_{\rm acc} \mathcal{E}_{\rm SF}^{-1}\,; \\
 \label{eq:metal:approx}
 \frac{\mathcal{Y}}{Z}   & \approx & \epsilon_{\rm acc} \mathcal{E}_{\rm SF}^{-1} \,.
\end{eqnarray}
In this case, the star formation efficiency ($\mathcal{E}_{\rm SF}$)
and the chemical evolution is completely determined by the gas exchange 
between the galaxies and their environment, independent of the gas 
content of the galaxy. The choice of star formation law is also not important unless 
it gives a gas mass that is comparable to $f_{\rm b}M_{\rm h}$.
This set of equations is basically equivalent to equations (16) and (18) 
in \citet{Dave12}, which are derived directly from the assumption that gas inflow, 
outflow and consumption by star formation are in equilibrium. This approximate  
model was adopted by \citet{Henry13} to evaluate the plausibility 
of different wind models (and models with no wind). We caution, 
however, that this simplified model is not general, and is only valid under the 
assumptions described above.   

\subsection{Models with strong gas outflow}

A commonly adopted assumption in galaxy formation models is that 
halos accrete baryons at the maximum rate, $f_{\rm b}{\dot M}_{\rm h}$.
For halos with mass below $10^{12}\Msun$, where the radiative cooling 
timescale is always shorter than halo dynamical time, the gas accretion onto the 
central galaxy is also expected to follow the halo accretion.
We test the consequence of this basic assumption using our constrained model. 

Setting $\epsilon_{\rm acc}=1$, i.e. assuming galaxies are accreting
at the maximum rate, we can calculate the net yield $\mathcal{Y}$ and the mass loading factor
$\epsilon_{\rm loss}$ using Eqs.\,(\ref{eq:gas:03}) and (\ref{eq:metal:03}).  
The results are shown as the red curves in Figure~\ref{flow_KSv1} for the 
Kennicutt-Schmidt star formation model
and in Figure\,\ref{flow_KMv1} for the Krumholz model, respectively.
Although the two star formation models lead to sizable differences in the 
gas mass, the predicted mass loading factors and net yields are very similar, 
suggesting that this uncertainty does not strongly affect the estimates of 
the yield and mass loading factor. The reason for this is that the conditions 
leading to the approximate model given by  Eqs.\,(\ref{eq:gas:approx}) 
and \,(\ref{eq:metal:approx}) are valid, so that 
${\cal E}_{\rm SF}$ and ${\cal Y}$ are independent of $M_{\rm g}$. 
In this case the gas exchange between the galaxy and 
the environments is rapid. For example, the required loading 
factor for $10^{11}\Msun$ halos can be as high as 10 to 20.

With the use of the fiducial $M$-$Z$ relations as constraints,
the net yield $\mathcal{Y}$ predicted 
exceeds the intrinsic yield, $y$ (shown as 
green horizontal lines in the upper panels of Figs.\,\ref{flow_KSv1} and \ref{flow_KMv1}) 
at least since $z\approx2$. The value of $\mathcal{Y}$ defined above is related to a number 
of factors:
(i) the intrinsic yield $y$; 
(ii) the value of $Z_{\rm w}$ which is determined by 
how well the metals produced by stars are mixed with the ISM; 
and 
(iii) the value of $Z_{\rm r}$ which is determined by  
     the history of the galaxies.
In general, the value of $\mathcal{Y}$ cannot exceed
that of $y$ because metals in both inflow and outflow must 
have been diluted. In large scale cosmological 
simulations \citep[e.g.][]{Dave12} and semi-analytic model 
of galaxy evolution \citep[e.g.][]{LY13}, metals produced by 
stars are assumed to be fully mixed with the ISM, so that
$Z_{\rm w} = Z$ is expected. Also, the observed metallicity 
of the ISM generally increases monotonically with time, 
so that $Z_{\rm r}<Z$. Putting all these together implies
$\mathcal{Y} = y+\epsilon_{\rm r}\left( Z_{\rm r}-Z \right) < y$.
On the other hand, in the case of no wind recycling, as assumed in \citet{Lilly13},
$\mathcal{Y} = y-\epsilon_{\rm r}\left(Z_{\rm w}-Z\right) \le y$.
Generally, as long as the recycled material is less enriched than
the wind, $\mathcal{Y}$ should always be no larger than the 
intrinsic yield $y$. Thus, under the assumption that 
gas accretion follows the accretion of the host dark halos,
the gas outflow would be required to be under enriched in metals
than what is to be expected, suggesting 
that the assumption $\epsilon_{\rm acc}=1$ is invalid. 

\begin{figure}
 \centering
 \includegraphics[width=0.9\linewidth]{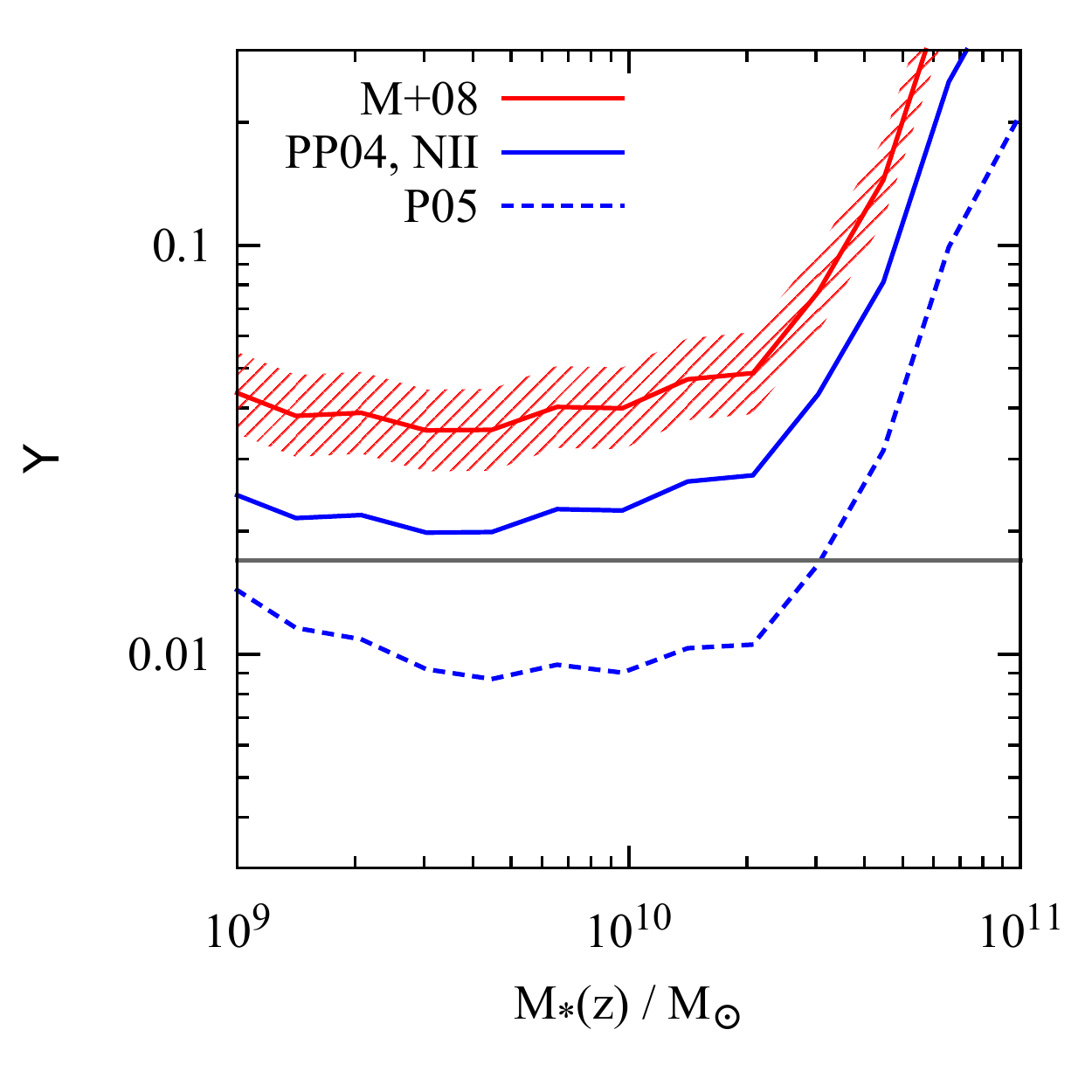}
 \caption{The net yield at $z=0$ calculated using Eq.\,(\ref{eq:metal:approx}) assuming $\epsilon_{\rm acc}=1$.
          The red solid line is based on the fiducial $M$-$Z$ relation from \citet{Maiolino08}
          and the band indicates the systematic uncertainty from the photoionization
          modeling. The blue solid line and the blue dashed line are based on the 
          metallicity measurements using the calibrations from \citet{PP04} and \citet{P05}
          respectively, both are based on $T_{\rm e}$ method.
          The horizontal line shows the intrinsic yield.}
 \label{uncertainty}
\end{figure}

As shown in Eq.\,(\ref{eq:metal:approx}), the net yield $\mathcal{Y}$ is 
roughly proportional to the gas phase metallicity measured $Z$. 
This provides a simple way to understand the 
systematic effects in the measured gas phase metallicity.
These effects are carefully analyzed in \citet{Kewley08}. 
The variance between different measurements 
using the photoionization modeling 
(the second method briefly described in \S\ref{constraints})
is about $0.2\,{\rm dex}$, and the resultant uncertainty in the 
net yields is shown as the red band shown in Figure~\ref{uncertainty}. 
It is clear that the net yield $\mathcal{Y}$ required 
is always larger than the intrinsic value $y$, in conflict with
the expectation that $\mathcal{Y} < y$. 
   
We have also used the $M$-$Z$ relations derived from 
the $T_{\rm e}$ method (or calibrations based on this 
method) to estimate the net yield, and the results 
are shown as the blue lines in Figure~\ref{uncertainty}. 
For HII regions with $12+\log_{10}({\rm O/H}) > 8.6$ the metallicity 
derived from this method is systematically lower. 
In particular, the $M$-$Z$ relation from \citet{P05} is 
lower by $0.7\,{\rm dex}$. 
The values of ${\mathcal{Y}}$ so derived  
are consistent with the intrinsic yield from stellar 
evolutions models, except at the massive end.
Unfortunately, this agreement cannot be taken seriously, 
because theoretical investigations have demonstrated that 
the $T_{\rm e}$-based methods tend to underestimate 
the metallicity in metal rich HII regions \citep[e.g.][]{Stasinska05}. 
What is clear, though, is that accurate 
measurements of the gas phase metallicity can provide stringent 
constraints on galactic inflow and outflow.

\subsection{Constraining gas inflow and outflow}

As discussed in the previous subsection, the natural assumption 
that the net yield ought to be lower than the intrinsic yield
requires a reduced rate for gas exchange between 
galaxy and its ambient medium.  
If the baryon mass exchange is too rapid via inflow of pristine gas or outflow of metal enriched ISM, 
the predicted gas phase metallicity would be too low when a reasonable value is assumed 
for the net yield. 
What this means is that we can constrain the upper limit for the inflow and outflow efficiencies, 
$\epsilon_{\rm acc}$ and $\epsilon_{\rm loss}$, by setting $\mathcal{Y}$ to 
its upper limit, namely setting $\mathcal{Y}=y$.
It can be shown that, as long as $Z_{\rm w} \ge Z$ 
and $Z_{\rm w}\ge Z_{\rm r}$, the relation 
$\mathcal{Y} = y$ requires both
$Z_{\rm w} = Z$ and $Z_{\rm w} = Z_{\rm r}$.
As mentioned above,  $Z_{\rm w} = Z$ implies
that the metals produced from star formation is 
fully mixed with the ISM. In this case, gas outflow is 
the least efficient in carrying metals out of galaxies.
The second condition, $Z_{\rm w} = Z_{\rm r}$, 
implies that some of the ejected gas is recycled
instantaneously while the rest is permanently lost.
The upper limits to $\epsilon_{\rm acc}$ so obtained 
are shown as the green lines in the lower panels 
of Figs.~\ref{flow_KSv1} and \ref{flow_KMv1}, 
while the upper limits to $\epsilon_{\rm loss}$ are
shown as the green lines in the middle panels 
of the same figures.  

The two different star formation models produce similar results. 
The variation in the gas content does not cause much variation 
in the estimate of $\epsilon_{\rm acc}$.    
The low star formation efficiency in low-mass galaxies 
is due to strong outflow at $z=2$, and to inefficient accretion at $z=0$. 
At $z=2$, the mass loading is roughly proportional to $M_{\rm h}^{-1}$ 
and is about $10$ for $10^{11}\Msun$ halos. 
At $z=0$ the mass loading depends only weakly on halo mass, 
with values close to $1$. Both the accretion efficiency, 
$\epsilon_{\rm acc}$, and the effective wind loading 
factor, $\epsilon_{\rm loss}$, drop by a 
factor of $2$ from $z=2$ to $z=0$. These drops  
are direct results of the evolution in the observed 
$M$-$Z$ relation, as our model assumes full mixing.  
At high redshift, a large fraction of metals are required to be lost with 
the ejected ISM in order to reproduce the relatively low metallicity, 
while at low redshift most of the metals are retained so as to
reproduce the increased metallicity. 

We also consider another special case in which 
there is no wind recycling, i.e. 
$\epsilon_{\rm loss} \to 0$.
\footnote{As shown in the following subsection, 
outflow of metals is always required to ensure  $\mathcal{Y}\le y$, and so  
strictly speaking $\epsilon_{\rm loss}$ cannot be exactly zero.} 
In this case, the accretion efficiency $\epsilon_{\rm acc}$ 
is required to be much lower than unity in order to maintain the 
total amount of cold gas in the disk.  
In this limit, the approximations given by Eqs.\,(\ref{eq:gas:approx})
and \,(\ref{eq:metal:approx}) are not valid anymore, and  
the derived $\epsilon_{\rm acc}$ depends on the star formation 
models adopted. For instance, since the gas fraction at high $z$ predicted by 
the Krumholz model is systematically higher than 
the prediction of the Kennicutt-Schmidt model,
the required gas accretion at low redshift is much lower,
because the star formation at low redshift can be fueled 
by the gas accumulated earlier in the galaxy.
If reincorporation of ejected gas is taken into 
account, it is possible that $\epsilon_{\rm loss} < 0$.
In this case, the corresponding $\epsilon_{\rm acc}$ and 
$\mathcal{Y}$ will occupy the grey hatched areas shown 
Figs.~\ref{flow_KSv1} and \ref{flow_KMv1}.

The boundaries we draw are based on the fiducial $M$-$Z$ relations and 
the fiducial intrinsic oxygen yield. As mentioned 
above, the systematic uncertainty in the metallicity estimate 
using detailed photoionization modeling is 
$\pm 0.1\,{\rm dex}$ around the mean. 
Since in the full mixing model, which gives
the upper limits of $\epsilon_{\rm acc}$ and $\epsilon_{\rm loss}$, 
the simple proportionality in Eq.\,(\ref{eq:metal:approx}) holds, 
a change by $\pm 0.1\,{\rm dex}$ in metallicity simply leads
to a change of $\mp 0.1\,{\rm dex}$ in $\epsilon_{\rm acc}$ and  
to a change of $\pm 0.1\,{\rm dex}$ in $\epsilon_{\rm loss}$.
However, if the $T_{\rm e}$-based metallicity is used as 
model constraint, the assumption that gas accretion 
into the galaxy follows the accretion of dark matter,
i.e. $\epsilon_{\rm acc}\approx 1$, is still permitted by the 
metallicity measurements, as shown in Figure\,\ref{uncertainty}. 

\begin{figure*}
 \centering
 \includegraphics[width=0.8\linewidth]{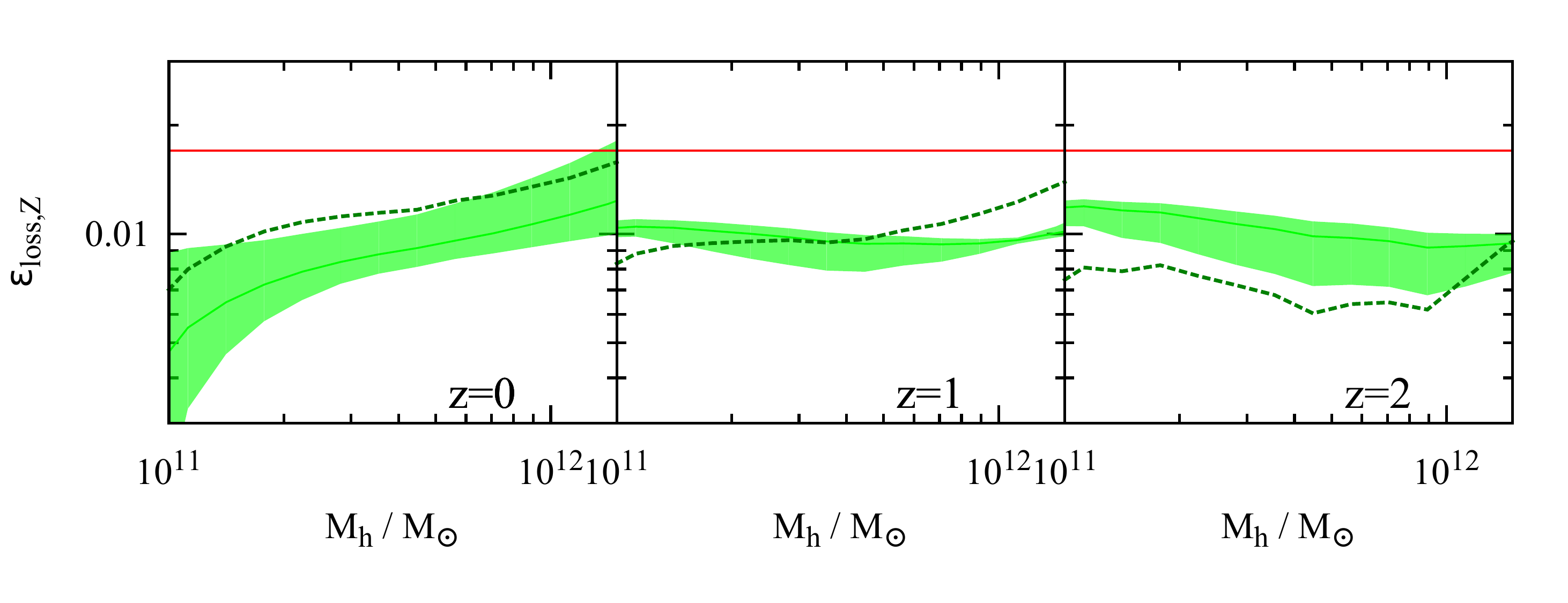}
 \caption{The loading factor of metal loss rate.
          Light green represents the Kennicutt-Schmidt star formation law
          and dark green represents the Krumholz star formation model.
          The curves are the predictions of the best fitting model in
          Table\,\ref{SFL_param} and the bands are obtained by marginalizing
          the uncertainties in the parameters.
          The red horizontal lines are the intrinsic yield of oxygen.}
 \label{flow_Z}
\end{figure*}

The oxygen yield from \citet{Kobayashi06} is about $0.01$. 
This number is quite close to the blue lines in the upper panels of 
Figs.\,\ref{flow_KSv1} and \ref{flow_KMv1},
which are obtained by setting $\epsilon_{\rm loss} = 0$.
This suggests that the combination of the \citet{Kobayashi06} 
chemical evolution model with the metallicity measurements 
using detailed photoionization modeling strongly prefers 
a weak outflow scenario, even at $z\approx 2$.

\subsection{Metal loss}

The loading factor of metal loss rate $\epsilon_{\rm loss, Z}$
can be directly estimated from Eq.\,(\ref{eq:metal:02}), and
the estimate is independent of the rates of gas inflow and outflow.
Figure~\ref{flow_Z} shows $\epsilon_{\rm loss, Z}$ as a function 
of halo mass at three different redshifts. 
As one can see, net metal outflow is always required, 
i.e. $\epsilon_{\rm loss, Z}>0$, for different halos 
at different redshifts, regardless of the gas outflow.
The loading factor predicted with the Kennicutt-Schmidt 
law is about $0.01$, which is about $60\%$ of the yield
(indicated by the horizontal lines), and depends only 
weakly on redshift and the mass of the host halos.
This is consistent with the finding of \citet{Peeples14},
that is about $75\%$ of the metals ever produced do not stay
in the host galaxies.
The prediction using the Krumholz model is similar 
except that at $z=2$ the mass loading factor is lower. 
The reason for this difference is that the Krumholz 
model predicts higher cold gas fraction at $z=2$, 
and so a larger fraction of newly produced metals can be 
stored in the ISM instead of going out with the wind.

\section{Conclusions and Implications}
\label{conclusion}

\begin{figure}
 \centering
 \includegraphics[width=0.9\linewidth]{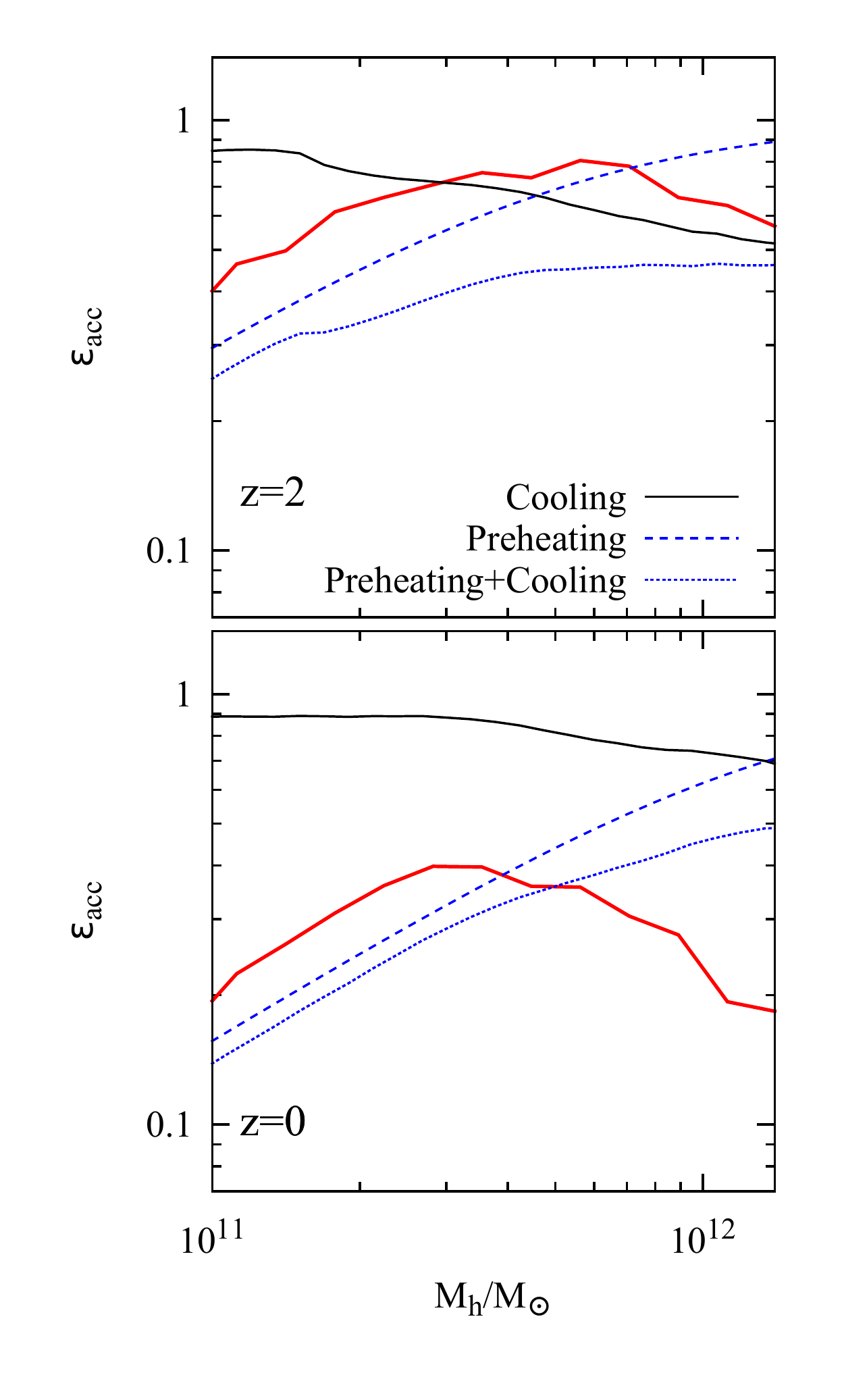}
 \caption{Comparison between the upper limit of the accretion efficiency
          derived from our empirical model (red lines) with 
          a number of physical models.
          The black solid lines are cooling efficiency of halo gas \citep{Croton06}.
          The blue dashed lines are the efficiency of accretion into dark matter halos
          in the preheating model of \citet{LY14} and the blue dotted lines
          show the accretion to the galaxies in the same model.}
 \label{phys}
\end{figure}

In the present paper, we have combined 
up-to-date observational constraints, including the 
star formation - halo mass relations \citep{LZ14b}, the gas 
phase metallicity - stellar mass relations \citep{Maiolino08},
and the gas mass fraction of local galaxies \citep{Peeples11},
and used a generic model  to investigate how the contents, inflow and outflow 
of gas and metals evolve in the ecosystem of a low-mass galaxy. 
The goal is to understand the 
underlying physics responsible for the low star formation 
efficiency in halos with masses between 
$10^{11}\Msun$ and $10^{12}\Msun$.
Our conclusions are summarized in the following. 

We adopt both the Kennicutt-Schmidt and the Krumholz 
models of star formation and combine each of them with 
the star formation histories of galaxies derived from the empirical model of  
\citet{LZ14b} to constrain the gas contents in galaxies 
up to $z=2$. We find that
(i) 
The gas mass to stellar mass ratio in general 
increases with redshift because of the increase of SFR;
(ii) 
The Krumholz model predicts a higher gas mass fraction 
at high redshift than the Kennicutt-Schmidt model, 
especially in dwarf galaxies, because of its dependence 
on metallicity and because of the metallicity evolution of the ISM;
(iii)
The Krumholz model predicts that the ISM of galaxies 
is dominated by the molecular gas at $z>1$, 
with the molecular gas to stellar mass ratio
increasing from $\sim 0.1$ at $z=0$ to $\sim 1$ at $z=2$;
(iv)
The baryon mass ratio, $(M_{\rm g}+M_{\rm \star})/M_{\rm h}$, 
is, since $z=2$,  always much less than the universal baryon mass fraction. 

Using the gas mass estimated from the star formation laws
together with other observational data,  we derive constraints 
on the gas inflow and outflow rates. Independent of the gas 
outflow rate, metal outflow is always required at different redshift. 
The metal mass loading factor is about $0.01$,
or about $60\%$ of the metal yield, and this factor depends 
only weakly on halo mass and redshift.

In spite of the degeneracy between gas inflow and outflow, and 
the uncertainties in modeling how metals are mixed with the medium, 
we can still put constraints on gas inflow and outflow. 
As the galactic wind material is expected to be more metal enriched 
than both the ISM and the material ejected at an earlier epoch, we can derive 
stringent upper limits on the accretion rate of primordial gas 
and on the net gas mass loss rate in the outflow.  We find that
(i) 
At $z\sim 0$, the low star formation efficiency is mainly caused by
the low accretion rate. The maximum loading factor of the mass 
loss is about one while the maximum accretion efficiency factor 
[${\dot M}_{\rm acc}/(f_{\rm b}{\dot M}_{\rm h})$] is between
$0.3$ and $0.4$;
(ii) 
At $z\sim 2$, strong gas mass loss is allowed. 
The maximum loading factor allowed by the observational constraints 
is about 10 for $10^{11}\Msun$ halos, and is inversely 
proportional to halo mass. These upper limits do not depend 
significantly on the star formation laws adopted, 
because the exact amount of gas in the galaxies is irrelevant 
in estimating the rate of gas exchange, as long as 
$M_{\rm g} \ll f_{\rm b}M_{\rm h}$, which is roughly the case
based on our model inferences.

In a typical semi-analytic model of galaxy formation, 
the mass accretion into a {\it halo} is usually 
assumed to be $f_{\rm b}\dot{M}_{\rm h}$, and 
the mass accretion into the {\it central galaxy}
is determined by the cooling rate of the gaseous halo. 
Following the cooling model of \citet{Croton06} (also in \citealt{LY11})
and assuming a metallicity of $0.1Z_{\odot}$ in the coronal gas, 
we calculate the efficiency of mass accretion in 
such a process, and the value of  $\epsilon_{\rm cool} \equiv 
\dot{M}_{\rm cool}/(f_{\rm b}{\dot M}_{\rm h})$ is
shown as the black solid lines in Figure~\ref{phys}. 
We see that $\epsilon_{\rm cool}\sim 1$ for a $10^{11}\Msun$ 
halo, and is roughly proportional to $M_{\rm h}^{-0.2}$ in the halo 
mass range shown in the figure. At $z=2$ $\epsilon_{\rm cool}$
is close to the upper limit of $\epsilon_{\rm acc}$ we have 
derived, but at $z=0$ it is significantly larger. 
The discrepancy between this prediction 
and our empirically derived constraint 
suggests that either accretion of the IGM into dark
halos must be reduced or the cooling of the halo gas 
must be slowed down.

The scenario of galaxy formation in a preheated medium
was first proposed in \citet{Mo02} in order to explain 
the observed stellar mass functions and HI mass function.
\citet{LY14} suggested that the extended gas disks provide
independent supports to such a scenario.
They considered an ``isentropic'' accretion model, in which 
the IGM is assumed to be preheated to a certain level
at $z<2$ so that the gas accretion rate into low-mass halos 
is reduced. The hot gaseous halos formed in this way are 
less concentrated and cooling can happen even 
in the outer part of a halo, where the specific angular 
momentum is higher, producing a disk size - stellar 
mass relation that matches observation.
Using the entropy model explored in \citet{LY14} we 
have calculated the accretion efficiency of the pre-heated 
IGM into {\it dark matter halos}, which is shown as the 
blue dashed lines in Figure~\ref{phys}. The corresponding 
accretion rate of the central galaxies due 
to radiative cooling of the gaseous halos is shown 
as the blue dotted lines. At $z=2$, the accretion 
efficiency lies below the upper limit, and so the model
is compatible with our results. 
At $z=0$, the predicted accretion efficiency is 
consistent with the upper limit we obtained
for halos with masses below $4\times10^{11}\Msun$
but is higher by a factor of $\sim 2$ for Milky Way mass halos.

It is still unclear how the IGM is preheated. In addition to 
the possibilities listed in \S\ref{intro},
\citet{LZ14c} proposed that intermediate mass central
black holes can serve as a promising source.
According to \citet{LZ14c}, such black holes form 
from the major merger between dwarf galaxies at $z > 2$ 
and is able to heat the surrounding
IGM to a entropy tested in \citet{LY14}.

For Milky Way mass halos, preventing the IGM from 
collapsing with the dark matter requires an entropy level
that is much larger than what \citet{LY14} suggests. 
Such a high level of preheating may over-quench 
star formation in smaller galaxies. It is more likely 
that some other {\it preventive} (rather than ejective) mechanisms 
may reduce {\it gas cooling} in such galaxies at low redshift, 
instead of preventing gas accretion into the {\it host halo}.  
For example, a central black hole may keep halo gas hot via 
the ``radio mode" feedback, preventing it from further 
cooling \citep{Croton06}.  Clearly, it is important to 
examine if such ``radio mode"  feedback is also 
operating in Milky Way size galaxies, or other processes 
have to be invoked. Using hydrodynamic 
simulation of Milky Way mass galaxies, \citet{Kannan14} 
found that the ionizing photons from local young and 
aging stars can effectively reduce the cooling of the halo gas, 
which may provide another promising preventive mechanism to 
reduce star formation efficiency in such galaxies at low
redshift, as predicted by our empirical model

\section*{Acknowledgements}

We thank Frank van den Bosch for helpful comments.
HJM would like to acknowledge the support of NSF AST-1109354.

\end{document}